\documentclass[preprint]{aastex}%
\usepackage{pstricks}
\makeindex

\def\lsim{\mathrel{\rlap{\lower4pt\hbox{\hskip1pt$\sim$}}
    \raise1pt\hbox{$<$}}}                
\def\gsim{\mathrel{\rlap{\lower4pt\hbox{\hskip1pt$\sim$}}
    \raise1pt\hbox{$>$}}}                

\def\be{\begin{equation}}
  \def\ee{\end{equation}}
\def\bea{\begin{eqnarray}}
  \def\eea{\end{eqnarray}}
\begin{document}

\title{The Cosmic Microwave Background Bipolar Power Spectrum: 
\\ Basic Formalism and Applications} 
\author{Amir Hajian\altaffilmark{1} and Tarun Souradeep\altaffilmark{2}} 
\affil{Inter-University Centre for   Astronomy and Astrophysics,\\ Post Bag 4, Ganeshkhind, Pune 411007,  India} 
\altaffiltext{1}{amir@iucaa.ernet.in}
\altaffiltext{2}{tarun@iucaa.ernet.in}

\begin{abstract}                   

We study the statistical isotropy (SI) of temperature fluctuations of the CMB as distinct from Gaussianity. We present a detailed formalism of the bipolar power spectrum (BiPS) which was introduced as a fast method of measuring the statistical isotropy by Hajian \& Souradeep 2003. The method exploits the existence of patterns in the real space correlations of the CMB temperature field. We discuss the applications of BiPS in constraining the topology of the universe and other theoretical scenarios of SI violation.  Unlike the traditional methods of search for cosmic topology, this method is computationally fast. We also show that BiPS is potentially a good tool to detect the effect of observational artifacts in a CMB map such as non-circular beam, anisotropic noise , etc. Our method has been successfully applied to the Wilkinson Microwave Anisotropy Probe sky maps by Hajian et al. 2004, but no strong evidence of SI violation was found. 
\end{abstract}
\section{Introduction}
The first-year {\it Wilkinson Microwave Anisotropy Probe} ({\it WMAP}) observations  are consistent with predictions of the concordance  $\Lambda$CDM model with scale-invariant and adiabatic fluctuations which have been generated during the inflationary epoch \cite{hin_wmap03, kogut_wmap03, sper_wmap03, page_wmap03, peiris_wmap03}. After the first year of {\it WMAP} data, the SI of the CMB anisotropy ({\it i.e.} rotational invariance of n-point correlations) has attracted considerable attention.  Tantalizing evidence of SI breakdown (albeit, in very different guises) has mounted in the {\it WMAP} first year sky maps, using a variety of different statistics. It was pointed out that the suppression of power in the quadrupole and octopole are aligned \cite{maxwmap}.  Further ``multipole-vector'' directions associated with these multipoles (and some other low multipoles as well) appear to be anomalously correlated \cite{cop04, schw04}. There are indications of asymmetry in the power spectrum at low multipoles in opposite hemispheres \cite{erik04a, han04, nas04}. Possibly related, are the results of tests of Gaussianity that show asymmetry in the amplitude of the measured genus amplitude (at about $2$ to $3\sigma$ significance) between the north and south galactic hemispheres \cite{par04, erik04b, erik04c}. Analysis of the distribution of extrema in {\it WMAP} sky maps has indicated non-gaussianity, and to some extent, violation of SI \cite{lar_wan04}. However, what is missing is a common, well defined, mathematical language to quantify SI (as distinct from non Gaussianity) and the ability to ascribe statistical significance to the anomalies unambiguously.

In order to extract more information from the rich source of information provided by present (and future) CMB maps, it is important to design as many independent statistical methods as possible to study deviations from standard statistics, such as statistical isotropy (SI). Since SI can be violated in many different ways, various statistical methods can come to different conclusions. Because each method by design is more sensitive to a special kind of SI violation. 

As a statistical tool of searching for departures from SI, we use the bipolar power spectrum (BiPS) which is sensitive to structures and patterns in the underlying total two-point correlation function \cite{us_apjl, us_pascos}. The BiPS is particularly sensitive to real space correlation patterns (preferred directions, etc.) on characteristic angular scales. In harmonic space, the BiPS at multipole $\ell$ sums power in off-diagonal elements of the covariance matrix, $\langle a_{lm} a_{l'm'}\rangle$, in the same way that the `angular momentum' addition of states $l m$, $l' m'$ have non-zero overlap with a state with angular momentum $|l-l'|<\ell<l+l'$. Signatures, like $a_{lm}$ and $a_{l+n m}$ being correlated over a significant range $l$ are ideal targets for BiPS. These are typical of SI violation due to cosmic topology and the predicted BiPS  in these models have a strong spectral signature in the bipolar multipole $\ell$ space \cite{us_prl}. The orientation independence of BiPS is an advantage since one can obtain constraints on cosmic topology that do not depend on the unknown specific orientation of the pattern ({\it{e.g.}}, preferred directions). 

In the rest of this section we briefly introduce some basic statistical properties of CMB and discuss the statistical isotropy as distinct from Gaussianity. In sections 2 and 3 we introduce BiPS in real as well as harmonic space. In section 4 and 5 we define an unbiased estimator for BiPS and compute its cosmic variance in details. In section 6 we briefly introduce some possible sources of breakdown of SI. In section 7 we discuss the results of applying the method to the simulations and WMAP data. We draw our conclusions in section 8. For completeness we have also included details of calculations in 6 appendices to keep the paper easily readable. 

\subsection{Statistical Properties of the CMB Anisotropy}
The temperature anisotropies of the CMB are described by a random field,  $\Delta T(\hat{n})=T(\hat{n})-T_0$, on a 2-dimensional surface of a sphere (the so called surface of last scattering), where $\hat{n}=(\theta,\phi)$ is a unit vector on the sphere and $T_0=\int{\frac{d\Omega_{\hat{n}}}{4\pi}T(\hat{n})}$ represents the mean temperature of the CMB.  The statistical properties of this field can be characterized by the $n$-point correlation functions
\be \label{npoint}
\langle \Delta T(\hat{n}_1)\Delta T(\hat{n}_2)\cdots \Delta T(\hat{n}_n)\rangle.
\ee
Here the bracket denotes the ensemble average, {\it{i.e.}} an average over all possible configurations of the field. If the field is Gaussian in nature, the connected part of the $n$-point functions disappears for $n > 2$. The non-zero (even-$n$)-point correlation functions can be expressed in terms of the $2$-point correlation function. As a result, a Gaussian distribution is completely described by the two-point correlation function. 

The statistics of the temperature fluctuations is Gaussian because they are linearly connected to quantum vacuum fluctuations of a very weakly interacting scalar field (the inflaton). Perturbations in primordial gravitational potential, $\Phi_0$, generate the CMB anisotropy, $\Delta T$. The linear perturbation theory gives a linear relation between  $\Phi_0$ and  $\Delta T$.
$\Phi_0({\mathbf{x}})$ is assumed to be Gaussian, because in inflation, quantum fluctuations produce Gaussian random phase variations in the gravitational potential
\be
\Phi_0({\mathbf{x}})=\int{\frac{{\mathrm{d}}^3 {\mathbf{k}}}{(2\pi)^3}\Phi_0(\mathbf{k})\exp{i \mathbf{k}\cdot \mathbf{x}}}
\ee
with $\langle \Phi_0({\mathbf{k}}) \Phi_0^*({\mathbf{k'}})\rangle = P(k) \delta({\mathbf{k}}-{\mathbf{k'}})$, the power spectrum of potential fluctuations at that epoch.  In the linear temperature approximation, in a spatially flat universe, the temperature anisotropies at the surface of last scattering can be linearly related to the potential fluctuations at that epoch \footnote{This description is very simple but naive. It does not take the line of sight effects such as integrated Sachs-Wolfe effect into account. For a more careful description see Appendix \ref{AppGauss}.}
\be
\Delta T(\hat{n})=\int{\frac{{\mathrm{d}}^3 {\mathbf{k}}}{(2\pi)^3}e^{i {\mathbf{k}}\cdot \hat{n}\tau_{rec}}\Phi_0({\mathbf{k}})g_T(k)},
\ee
where $\tau_{rec}$ is the comoving conformal lookback time of the recombination epoch ($\tau$ is the conformal time $\tau=\int{\mathrm{d}t/a}$), $\phi(\tau)$ is a normalization constant for potential perturbations which is equal to unity today, and $g_T(k)$ is the linear radiation transfer function  which describes the relationship between potential fluctuations and temperature fluctuations at recombination. For small $k$, $g_T(k)=1/3$, which is known as the fact that for temperature fluctuations on super-horizon scales at the decoupling epoch, the Sachs-Wolfe effect \cite{SachsWolfe} dominates and hence on large angular scales
\be
\frac{\Delta T}{T} (\hat n) \, = \, \frac{1}{3} \Phi_{LS}(\hat n\tau_{rec},\tau_0)
\ee
On sub-horizon scales, $g_T$ oscillates (acoustic oscillation), and we need to evolve the coupled matter-radiation fluid, {\it i.e.} solve the Boltzmann photon transport equations coupled with the Einstein equations for $g_T$. Codes like CMBFAST \footnote{Available at http://www.cmbfast.org/} compute $g_T(k)$ for different cosmological models. 

If we expand the gravitational potential into a series in powers of a density enhancement
 \be
 \label{expansion}
 \Phi({\mathbf{x}},t)=\Phi^{(1)}+\Phi^{(2)}+\cdots
 \ee
 The linear term $\Phi^1=\Phi_0({\mathbf{x}})$ and is assumed to be Gaussian .Non-linearity in inflation makes  $\Phi$ weakly non-Gaussian and will lead to non-linear terms, such as $\Phi^{(2)}$, in eqn. (\ref{expansion}).  However, since potential fluctuations in the early universe are small, $\langle \Phi^2 \rangle \sim 10^{-9}$, these non-linear effects are usually viewed as undetectable \cite{Munshi95, Spergel98} (for a nice review on non-Gaussianity see \cite{Bartolo}).  Hence, here we limit our study to Gaussian CMB anisotropy field, where the two-point correlation function 
\be \label{2pointcorrelation}
C(\hat{n},\,\hat{n}^\prime)\equiv\langle\Delta T(\hat n)\Delta T(\hat n^\prime)\rangle
\ee 
contains all the statistical information encoded in the field. 

It is convenient to expand the temperature anisotropy field into spherical harmonics, the orthonormal basis on the sphere, as
\be \label{yelemexpand}
\Delta T(\hat{n}) \, = \, \sum_{l,m} a_{lm}Y_{lm}(\hat{n}) \,\,,
\ee
where the complex quantities, $a_{lm}$ are given by
\be \label{alm}
a_{lm} = \int{\mathrm{d} \Omega_{\hat{n}}Y_{lm}^{*}(\hat{n})
\Delta T(\hat{n})}.
\ee
For a Gaussian CMB anisotropy, $a_{lm}$ are Gaussian random variables and hence, the {\it covariance matrix},   $\langle a_{lm}a^*_{l^\prime m^\prime}\rangle$, fully describes the whole field. 

\subsection{Statistical Isotropy of Temperature Anisotropies}
A random field $\Phi({\mathbf{r}})$ is {\it statistically homogeneous} (in the wide sense) if the
mean and the second moment (covariance) of that remain invariant under the coordinate transformation 
$ {\bf r} \rightarrow {\bf r}\,+\,\delta {\bf r}$. That is to say,
\bea 
\langle {\Phi}( {\bf r}) \rangle \,&=&\, \langle \Phi({\bf r}\,+\,
\delta {\bf r})\rangle, \\ \nonumber
C_{\Phi}( {\bf r}_1 , \, {\bf r}_2) \,&=&\, C_{\Phi}( {\bf r}_1 +\,
\delta {\bf r} , \, {\bf r}_2 +\, \delta {\bf r}).
\eea
The first condition implies that  $\langle {\Phi} \rangle\, =\, $ const. Putting $\delta {\bf r}\, = \, -{\bf r}_2$
in the second condition we find that $C_{\Phi}( {\bf r}_1 , \, {\bf r}_2) \,=\, C_{\Phi}( {\bf r}_1-{\bf r}_2,\, 0)$, {\it i.e. } the covariance of a statistically homogeneous field is dependent only on the {\it difference} ${\bf r}_1 - {\bf r}_2$, and not on ${\bf r}_1$ and  ${\bf r}_2$ separately. So instead of $C_{\Phi}( {\bf r}_1-{\bf r}_2,\, 0)$ we can simply write $C_{\Phi}( {\bf r})$ which means:
\be
C_{\Phi}({\bf r})\,=\,\langle \tilde{\Phi}( {\bf r}_1) \tilde{\Phi}( {\bf r}_1+ {\bf r})\rangle.
\ee
where $ \tilde{\Phi} \equiv {\Phi} \,-\,\langle \Phi \rangle$.

Statistically homogeneous fields, with $C_{\Phi}({\bf r})$ dependent only on the {\it magnitude}(but not on the {\it direction}) of the vector $ {\bf r} \,=\, {\bf r}_2 \,-\, {\bf r}_1$ connecting the points ${\bf r}_1$ and ${\bf r}_2$
\be
\label{sigeneral}
C_{\Phi}({\bf r}) \,=\, C_{\Phi}(r), \,\,\,\,\,\,\,\,\,\,\, r\,=\,|{\bf r}|\,=\,\sqrt{{\bf r}\cdot{\bf r}},
\ee
are said to be {\it statistically isotropic}. The covariance of isotropic random fields is even and always real. 

For temperature anisotropies of the CMB, statistical isotropy is simply equivalent to {\it rotational invariance} of $n$-point correlation functions
\be \label{rotinv}
\langle \Delta T({\mathcal R}\hat{n}_1)\Delta T({\mathcal R}\hat{n}_2)\cdots \Delta T({\mathcal R}\hat{n}_n)\rangle = \langle \Delta T(\hat{n}_1)\Delta T(\hat{n}_2)\cdots \Delta T(\hat{n}_n)\rangle,
\ee
where $\Delta T({\mathcal R}\hat{n}_1)$ is the temperature anisotropy at ${\mathcal R}\hat{n}_1\,$ which is pixel  $\hat{n}_1\,$ rotated by the rotation matrix  ${\mathcal R}(\alpha,\beta,\gamma)$ for the Euler angles $\alpha,\beta$ and $\gamma$.
For two point correlations of CMB, where $|{\bf r_1}|=|{\hat{n_1}}\tau_{rec}|=$ const., the condition of statistical isotropy (eqn. \ref{sigeneral}) will translate into 
\be 
\label{statiso}
C(\hat{n},\,\hat{n}^\prime)\,=\, C(\theta), \,\,\,\,\,\,\,\,\,\,\,\,\,\, 
\theta\,=\,\arccos{(\hat{n} \cdot \hat{n}^\prime)}
\ee
which implies that the two point correlation is {\it just} a function of separation angle, $\theta$, between the two points on the sky. It is then convenient to expand it in terms of Legendre polynomials,
\be
\label{LegendreTransform}
C(\theta) \,=\, \frac{1}{4\pi}\sum_{l=2}^{\infty}(2l+1) C_l P_l(\cos{\theta}),
\ee
where $C_l$ is the widely used {\it angular power spectrum}.  The summation over $l$ starts from $2$ because the $l\,=\,0$ term is the monopole which in  the case of statistical isotropy the monopole is constant, and it can be subtracted out. The dipole $l\,=\,1$ is due to the local motion of the observer and is subtracted out as well. 
    
To obtain the statistical isotropy condition in harmonic space, we expand the left hand side of eqn. (\ref{rotinv}) in terms of spherical harmonics, each $\Delta T({\mathcal R}\hat{n}_i)$ may be expanded as 
\bea
\Delta T({\mathcal R}\hat{n}_1) &=& \sum_{l,m} a^{\mathcal R}_{lm} Y_{lm}({\mathcal R}\hat{n}_1) \\ \nonumber
&=& \sum_{l,m} a^{\mathcal R}_{lm} \sum_{m'} Y_{lm'}(\hat{n}_1) D^{l}_{m'm}({\mathcal R}),
\eea
in which $a^{\mathcal R}_{lm}$ are harmonic coefficients in rotated coordinates and $D^{l}_{m'm}({\mathcal R})$ is the finite rotation matrix in the $l m$-representation, Wigner $D$-function \cite{Var}. By substituting this into eqn.(\ref{rotinv}) and using the orthonormality law of spherical harmonics,  eqn.(\ref{A2}), we obtain the statistical isotropy condition in harmonic space
\be
\label{siharmonic}
\sum_{m'_1,m'_2\cdots,m'_n} \langle a^{\mathcal R}_{l_1m'_1} a^{\mathcal R}_{l_2m'_2}\cdots a^{\mathcal R}_{l_nm'_n} \rangle D^{l_1}_{m_1m'_1}({\mathcal R})D^{l_2}_{m_2m'_2}({\mathcal R})\cdots D^{l_n}_{m_nm'_n}({\mathcal R})= \langle a_{l_1m_1} a_{l_2m_2}\cdots a_{l_nm_n} \rangle.
\ee
For a Gaussian field, only the covariance, $\langle a_{lm}a^*_{l^\prime m^\prime}\rangle$, is important. As it is shown in Appendix \ref{moreharmonic}, the condition for statistical isotropy translates to a diagonal covariance matrix, 
\be
\label{diagonalcovmat}
\langle a_{lm}a^*_{l^\prime m^\prime}\rangle=C_{l} \delta_{ll^\prime}\delta_{mm^\prime}, 
\ee
which means that the angular power spectrum of CMB anisotropy, $C_l$, tells us everything we need to know about the (Gaussian) CMB anisotropy. When $\langle a_{lm} a^*_{l^\prime m^\prime}\rangle$ is not diagonal, the angular power spectrum $C_l$ does not have all the information of the field and one should also take the off-diagonal elements into account.
\subsection{Statistical Isotropy and Gaussianity}
Statistical isotropy and Gaussianity of CMB anisotropies are two independent and distinct concepts.  The Gaussianity of the primordial fluctuations is a key assumption of modern cosmology, motivated by simple models of inflation. And since the statistical properties of the primordial fluctuations are closely related to those of the cosmic microwave background (CMB) radiation anisotropy, a measurement of non-Gaussianity of the CMB will be a direct test of the inflation paradigm. If CMB anisotropy is Gaussian, then the two point correlation fully specifies the statistical properties. In the case of non-Gaussian fields, one must take the higher moments of the field into account in order to fully describe the whole field.  Statistical isotropy is a condition which guarantees that statistical properties of CMB sky (e.g. $n$-point correlations) are invariant under rotation. Hence, in a SI model, the iso-contours of two point correlations where one point is fixed and the other scans the whole sky, are concentric circles around the given point. Any breakdown of statistical isotropy will make these iso-contours non-circular. In addition to cosmological mechanisms, instrumental and environmental effects in observations easily produce spurious breakdown of statistical isotropy. 

A Gaussian field may or may not respect the statistical isotropy. In other words we can have
\begin{enumerate}
\item{Gaussian and SI models}: In these models two point correlation
  and $C_l$ have all the information of the field and  $C(\hat{n}_1,\hat{n}_2) = C(\hat{n}_1 \cdot \hat{n}_2)$
\item{Gaussian but not SI models}: two point correlation contains all the
  information of the field, but $C(\hat{n}_1,\hat{n}_2) \ne C(\hat{n}_1 \cdot \hat{n}_2)$. So $C_l$ is not adequate to describe the field and off-diagonal elements of covariance matrix should be taken into account.  
\item{Non-Gaussian but SI models}: two point correlation alone is not sufficient to describe the whole field and we need to take higher moments into account but every $n$-point correlation of the field is invariant under rotation.
\item{Non-Gaussian and non-SI models}: neither two point correlation has all the information nor it is invariant under rotation.  
\end{enumerate}
Here we confine ourselves to Gaussian fields where we  only need to consider  two point correlations. We will present a general way of describing fields of the kind (1) and (2) in the above list.

\section{The Bipolar Power Spectrum (BiPS)} 
Two point correlation of CMB anisotropies, $C(\hat{n}_1,\, \hat{n}_2)$, is a two point function on $S^2 \times S^2$, and hence can be expanded as
\be \label{bipolar}
C(\hat{n}_1,\, \hat{n}_2)\, =\, \sum_{l_1,l_2,L,M} A_{l_1l_2}^{\ell M}
\{Y_{l_1}(\hat{n}_1) \otimes Y_{l_2}(\hat{n}_2)\}_{\ell M},
\ee
where $A_{l_1l_2}^{\ell M}$ are   coefficients of the expansion (here after BipoSH coefficients) and  $\{Y_{l_1}(\hat{n}_1) \otimes Y_{l_2}(\hat{n}_2)\}_{\ell M}$ are the bipolar spherical harmonics which transform as a spherical harmonic with $\ell,\, M$ with respect to rotations \cite{Var} given by
\be 
\{Y_{l_1}(\hat{n}_1) \otimes Y_{l_2}(\hat{n}_2)\}_{\ell M} \,=\,
\sum_{m_1m_2} {\mathcal C}_{l_1m_1l_2m_2}^{\ell M} Y_{l_1 m_1}(\hat{n}_2)Y_{l_2 m_2}(\hat{n}_2),
\ee
in which ${\mathcal C}_{l_1m_1l_2m_2}^{\ell M}$ are Clebsch-Gordan coefficients. We can inverse-transform $C(\hat{n}_1,\, \hat{n}_2)$ to get the $A_{l_1l_2}^{\ell M}$ by multiplying both sides of eqn.(\ref{bipolar}) by $\{Y_{l'_1}(\hat{n}_1) \otimes Y_{l'_2}(\hat{n}_2)\}_{\ell'M'}^*$ and integrating over all angles, then the orthonormality of bipolar harmonics, eqn. (\ref{A2}), implies that
\be \label{alml1l2}
A_{l_1l_2}^{\ell M} \,=\,\int d\Omega_{\hat{n}_1}\int d\Omega_{\hat{n}_2} \,
C(\hat{n}_1,\, \hat{n}_2)\, \{Y_{l_1}(\hat{n}_1) \otimes Y_{l_2}(\hat{n}_2)\}_{\ell M}^*. 
\ee
The above expression  and the fact that $C(\hat{n}_1,\, \hat{n}_2)$ is symmetric under the exchange of $\hat{n}_1$ and $\hat{n}_2$ lead to the following symmetries of  $A_{l_1l_2}^{\ell M}$
\bea \label{sym}
A_{l_2l_1}^{\ell M}\,&=&\,(-1)^{(l_1+l_2-L)}A_{l_1l_2}^{\ell M}, \\ \nonumber
A_{ll}^{\ell M} \, &=& \, A_{ll}^{\ell M} \,\,\delta_{\ell,2k+1}, \,\,\,\,\,\,\,\,\,\,\,\,\,\,\,\,\,\,\,\,\, k=0,\,1,\,2,\,\cdots.
\eea

\begin{figure}[h]
  \includegraphics[scale=0.8, angle=0]{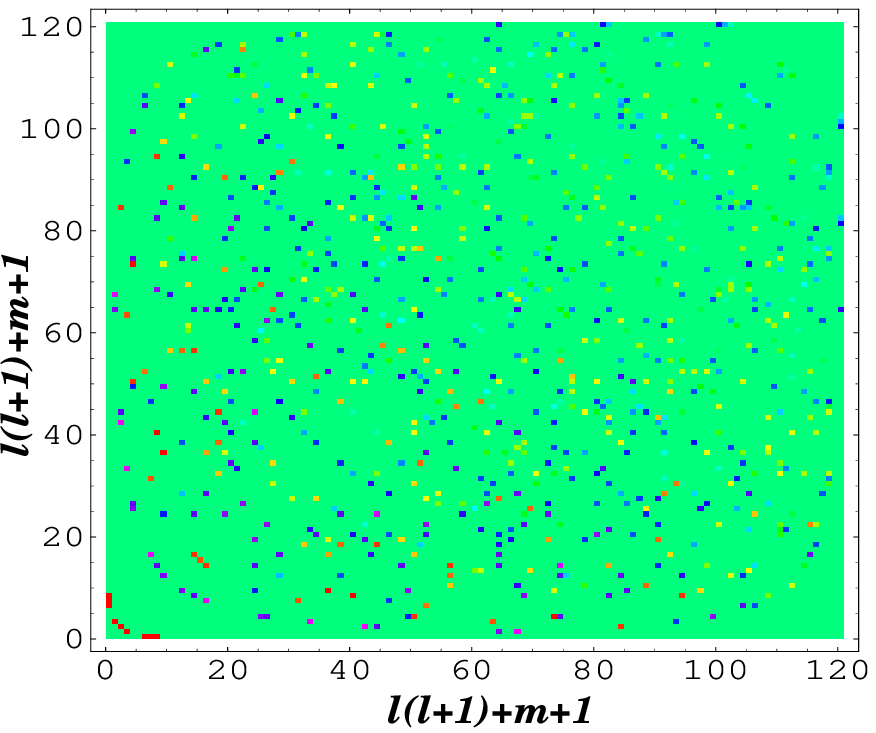}
  \includegraphics[scale=0.8, angle=0]{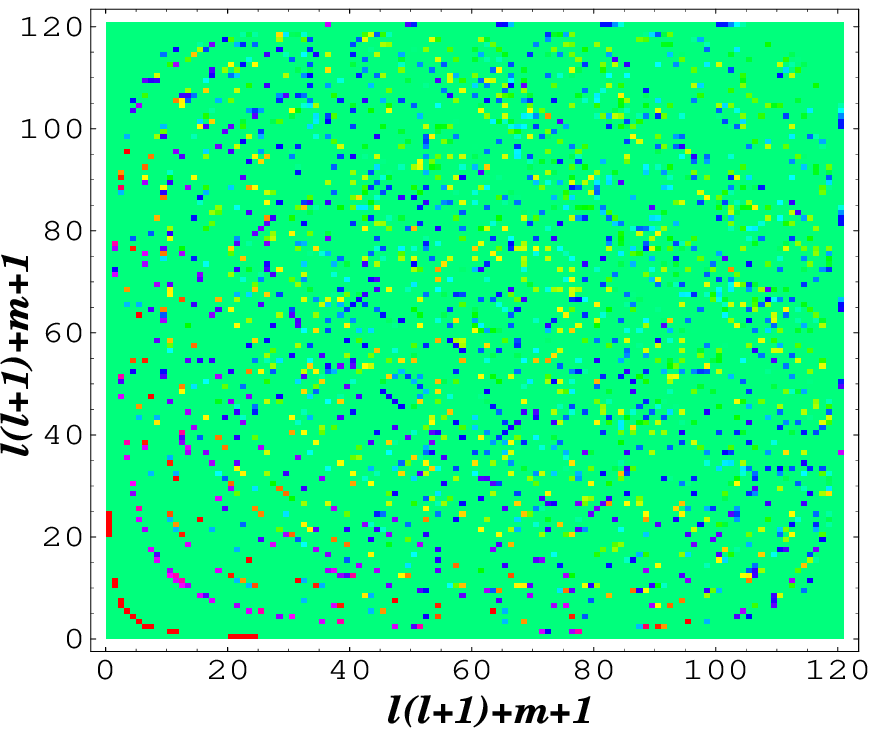}
  \caption{BipoSH coefficients are linear combinations of elements of the covariance matrix. Here $A^{2M}_{ll'}$ ({\it left}) and $A^{4M}_{ll'}$ ({\it right}) are plotted to show how BiPS covers the off-diagonal elements of the covariance matrix in harmonic space. }
  \label{ALM}
\end{figure}

We show in Appendix \ref{ALMproperties} that the Bipolar Spherical Harmonic (BipoSH) coefficients, $A_{l_1l_2}^{\ell M}$, are in fact linear combinations of off-diagonal elements of the covariance matrix,
\be \label{ALMvsalm}
A^{\ell M}_{l_1 l_2}\,=\, \sum_{m_1m_2} \langle a_{l_1m_1}a^{*}_{l_2 m_2}\rangle (-1)^{m_2} C^{\ell M}_{l_1m_1l_2 -m_2}.
\ee
This means that $A^{\ell M}_{l_1 l_2}$ completely represent the information of the covariance matrix. Fig. \ref{ALM} shows how  $A^{2 M}_{l_1 l_2}$ and  $A^{4 M}_{l_1 l_2}$ combine the elements of the covariance matrix.  When SI holds, the covariance matrix is diagonal and hence 
\bea \label{SIALM}
A_{ll^\prime}^{\ell M}\,&=&\,(-1)^l C_{l} (2l+1)^{1/2} \,  \delta_{ll^\prime}\, \delta_{\ell 0}\, \delta_{M0},
\\ \nonumber
A^{0 0}_{l_1 l_2}\,&=&\, (-1)^{l_1} \sqrt{2l_1+1}\, C_{l_1}\, \delta_{l_1l_2}.
\eea 
BipoSH expansion is the most general way of studying two point correlation functions of CMB anisotropy. The well known angular power spectrum, $C_l$ is in fact a subset of BipoSH coefficients.  

It is impossible to measure all $A^{\ell M}_{l_1 l_2}$ individually because of cosmic variance. Combining BipoSH coefficients into Bipolar Power Spectrum  reduces the cosmic variance\footnote{This is similar to combining $a_{lm}$ to construct the angular power spectrum, $C_l=\frac{1}{2l+1}\sum_{m}{|a_{lm}|^2}$, to reduce the cosmic variance}. BiPS of CMB anisotropy is defined as a convenient contraction of the BipoSH coefficients
\be \label{kappal}
\kappa_\ell \,=\, \sum_{l,l',M} |A_{ll'}^{\ell M}|^2 \geq 0.
\ee
The BiPS has interesting properties. It is orientation independent and is invariant under rotations of the sky. For models in which statistical isotropy is valid, BipoSH coefficients are given by eqn~(\ref{SIALM}). And results in a null BiPS, {\it i.e.}  $\kappa_\ell\,=\,0$ for every positive $\ell$, 
\be 
\kappa_\ell\,=\,\kappa_0 \delta_{\ell 0}.
\ee

So, non-zero components of BiPS imply the break down of statistical isotropy. And this introduces BiPS as a measure of statistical isotropy. 
\be
{\mathrm {\Huge STATISTICAL\,\,\,\, ISOTROPY}} \,\,\,\,\,\,\, \Longrightarrow \,\,\,\,\,\,\, 
\kappa_\ell\,=\,0 \,\,\,\,\,\,\, \forall \ell \ne 0.
\ee

\section{BiPS Representation in Real Space}
\begin{figure}[h]
  \begin{center}
    \includegraphics[scale=0.6]{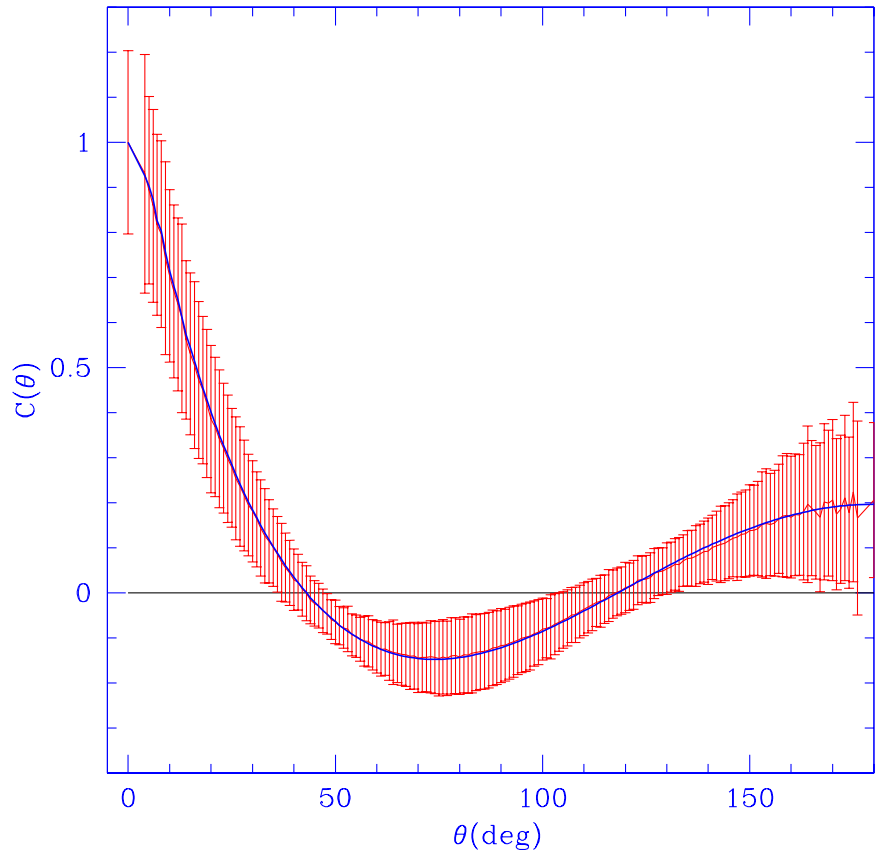}
  \end{center}
  \caption{$C(\theta)$ `recovered' from a map of a statistically isotropic model matches the original $C(\theta)$ shown as a solid line 
    from which we made the map. The error bars shown are computed from 10 realizations. Note that the size of error bar depends on the 
    size of $\theta$ bins.}
  \label{recov}
\end{figure}

Two point correlation of the CMB anisotropy is given by ensemble average over many universes (eqn.(\ref{2pointcorrelation})). But in reality, there is only one universe and the ensemble average is meaningless unless the CMB sky is SI, where the two point correlation function
$C(\theta)$ can be well estimated by the average product of temperature fluctuations in two directions $\hat n$ and $\hat n'$ whose angular separation is $\theta$ (see Fig. \ref{recov}),  this can be written as 
\begin{equation}
  \tilde C(\theta)\,= \int{\frac{\mathrm{d} \Omega_{\hat{n}}}{4\pi}\int{\frac{\mathrm{d} \Omega_{\hat{n}'}}{4\pi}\delta(\hat{n} \cdot \hat{n}'-\cos{\theta}) \Delta T(\hat n) \Delta T(\hat n')}}
  \label{bin_cth}
\end{equation}


If the statistical isotropy is violated the estimate of the correlation
function from a sky map given by a single temperature product 
\be
\tilde C(\hat{n}_1, \hat{n}_2)\,=\, \Delta T(\hat{n}_1) \Delta
T(\hat{n}_2) 
\ee 
is poorly determined.

Although it is not possible to estimate each element of the full correlation function $C(\hat{n}_1,\hat{n}_2)$, some measures of statistical isotropy of the CMB map can be estimated through suitably weighted angular averages of  $\Delta T(\hat{n}_1) \Delta T(\hat{n}_2)$. The angular averaging procedure should be such that the measure involves averaging over sufficient number of independent  measurements , but should ensure that the averaging does not erase all the signature of statistical anisotropy. Another important desirable property is that measure be independent of the overall orientation of the sky. Based on these considerations, we have proposed a set of measures of statistical isotropy~\cite{us_apjl} which can be shown that is equivalent to the  one in eqn.(\ref{kappal})
 \be \label{kl}
 \kappa^{\ell}\,=\, (2l+1)^2 \int d\Omega_{n_1}\int d\Omega_{n_2} \,
 [\frac{1}{8\pi^2}\int d{\mathcal R} \chi^{\ell}({\mathcal R})\, C({\mathcal R}\hat{n}_1,\, {\mathcal R}\hat{n}_2)]^2.
 \ee
 In the above expression, $ C({\mathcal R}\hat{n}_1,\, {\mathcal R}\hat{n}_2)$ is the two point correlation at 
 ${\mathcal R}\hat{n}_1\,$ and $ {\mathcal R}\hat{n}_2$ which are the coordinates of the two pixels 
 $\hat{n}_1\,$ and $\hat{n}_2$ after rotating the coordinate system through an angle $\omega$ where $(0\leq \omega \leq \pi)$
 about the axis ${\bf n}(\Theta, \Phi)$.  The direction of this rotation axis ${\bf n}$ is defined by the 
 polar angles $\Theta$ where $(0\leq \Theta \leq \pi)$ and $\Phi$, where $(0\leq \Phi \leq 2\pi)$. $\chi^{\ell}$ is the trace of the 
 finite rotation matrix in the $\ell M$-representation
 \be \label{chil} 
 \chi^{\ell}({\mathcal R})\,=\,\sum_{M=-\ell}^{\ell} D_{MM}^{\ell}({\mathcal R}),
 \ee
 which is called the {\it characteristic function}, or the character of the irreducible representation of rank $\ell$. It is invariant under rotations of the coordinate systems. Explicit forms of  $\chi^{\ell}({\mathcal R})$ are simple when ${\mathcal R}$
 is specified by $\omega,\, \Theta,\,\Phi$, then $\chi^{\ell}({\mathcal R})$ is completely determined by the rotation angle $\omega$ and 
 it is independent of the rotation axis ${\bf n}(\Theta, \Phi)$,
 \bea
 \chi^{\ell}({\mathcal R})\,&=&\,\chi^{\ell}(\omega)\,\\ \nonumber
 &=&\,\frac{\sin{[(2\ell+1)\omega/2]}}{\sin{[\omega/2]}}\,\,.
 \eea
 And finally $d{\mathcal R}$ in eqn(\ref{kl}) is the volume element of the three-dimensional rotation group and is given by
 \be 
 d{\mathcal R}\,=\, 4 \, \sin^2{\omega \over{2}}\, d\omega \, \sin{\Theta}\, d\Theta \, d\Phi\,\,.
 \ee
 As it is shown in Appendix (\ref{simplification}), this expression for BiPS may be simplified further as
\be
\label{simplifiedkl}
\kappa_{\ell}\, = \, \frac{2\ell+1}{8\pi^2} \int d\Omega_1 
\int d\Omega_2 C(\hat{q}_1,\hat{q}_2)  \int dR \chi_{\ell}(R) 
C(R\hat{q}_1,R\hat{q}_2)
\ee

For a statistically isotropic model  $ C(\hat{n}_1,\, \hat{n}_2)$ is invariant under rotation, and therefor  
 $ C({\mathcal R}\hat{n}_1,\, {\mathcal R}\hat{n}_2)\,=\,C(\hat{n}_1,\, \hat{n}_2)$. If we use this property to 
 substitute in eqn.(\ref{simplifiedkl}) for $ C({\mathcal R}\hat{n}_1,\, {\mathcal R}\hat{n}_2)$, and if we use the orthonormality of $\chi^{\ell}(\omega)$ (see eqn. \ref{A8}), we will recover the condition for SI,
 \be
 \kappa^{\ell} \, = \, \kappa^0 \delta_{\ell 0}.
 \ee
The proof of equivalence of BiPS definition in eqns.(\ref{kl}) and (\ref{kappal}) is as follows.  In eqn.(\ref{kl}), we substitute (\ref{chil}) for $\chi^{\ell}({\mathcal R})$ and from
\be 
C({\mathcal R}\hat{n}_1,\, {\mathcal R}\hat{n}_2)\,=\,\sum_{l_1l_2}^{\ell M'}A_{l_1l_2}^{\ell M'}
\sum_{M} D_{MM'}^{\ell}({\mathcal R}) \{Y_{l_1}(\hat{n}_1) \otimes Y_{l_2}(\hat{n}_2)\}_{\ell M},
\ee
we substitute for $ C({\mathcal R}\hat{n}_1,\, {\mathcal R}\hat{n}_2)$ and use the orthonormality of  $D_{MM}^{\ell}({\mathcal R})$ (see eqn.(\ref{A9})), we will obtain the eq.(\ref{kappal}). Real-space representation of BiPS is very suitable for analytical compution of BiPS for  theoretical models  where we know the analytical expression for the two point correlation of the model, such as theoretical models in \cite{us_prl}. This formalism would immensely simplify analytical calculations.  On the other hand, the harmonic representation of BiPS allows computationally rapid methods for BiPS estimation from a given CMB map. 

\section{Unbiased Estimator of BiPS}
\begin{figure}[p]
  \includegraphics[scale=0.3, angle=-90]{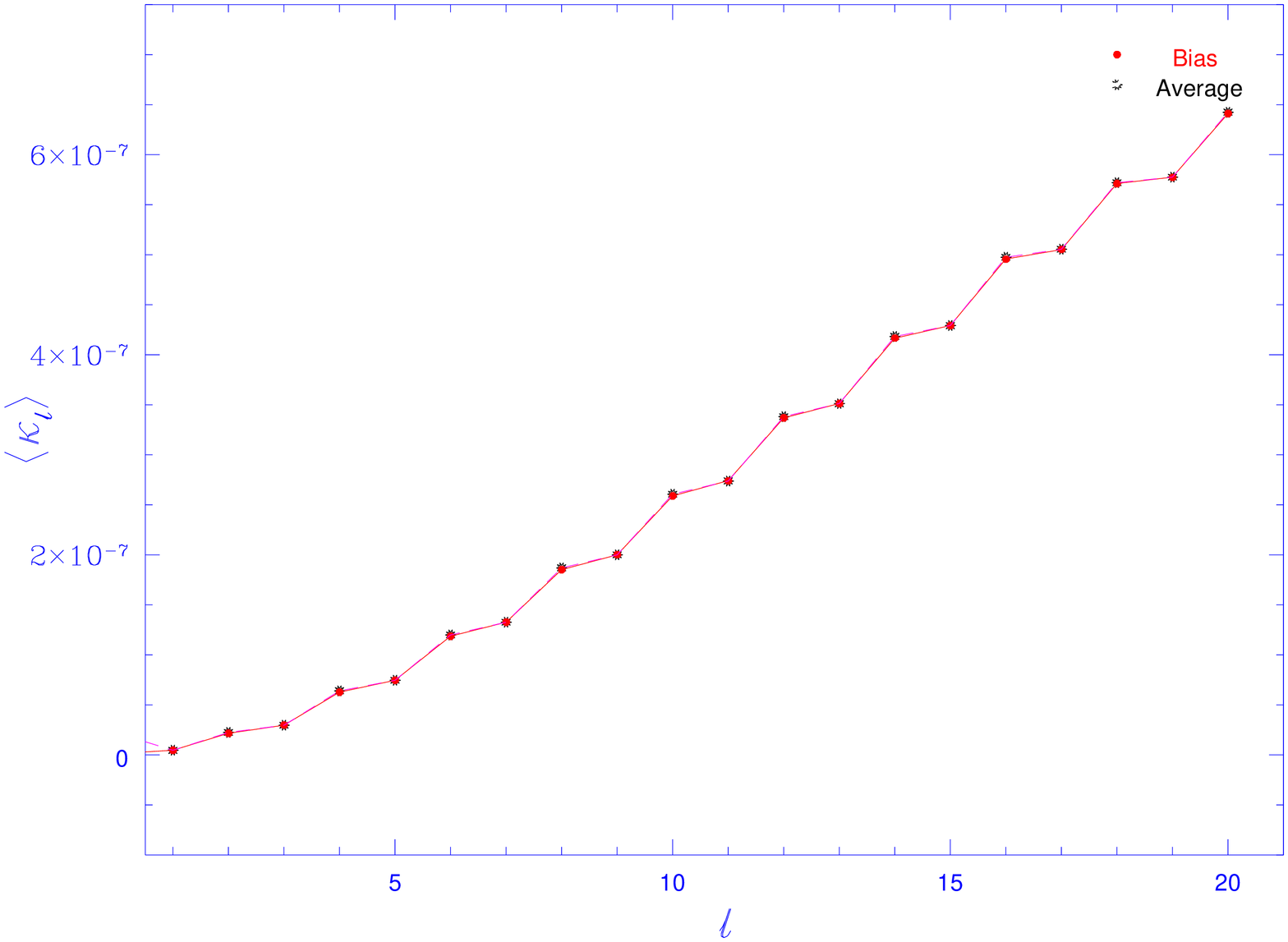}
  \includegraphics[scale=0.3, angle=-90]{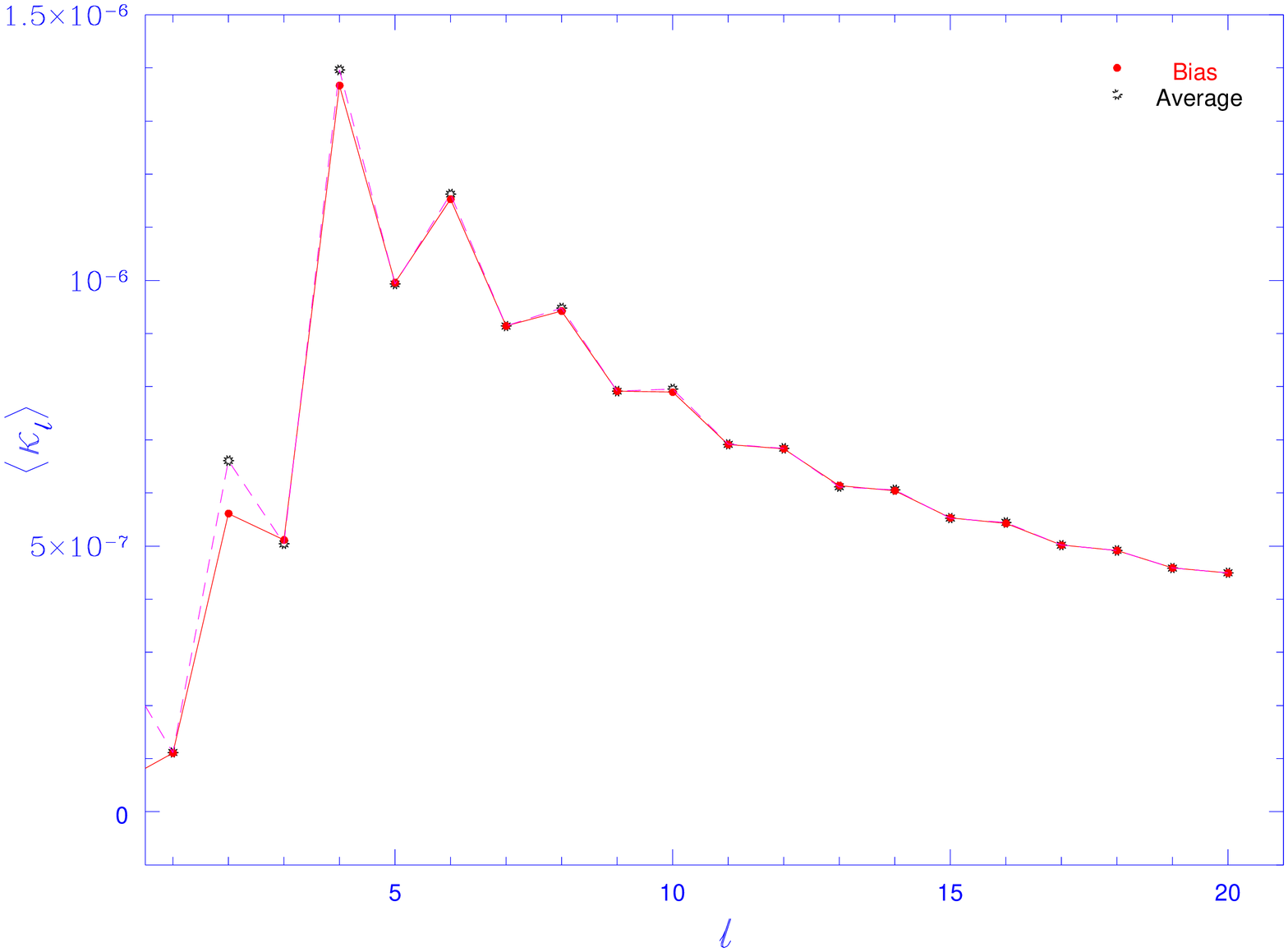}
  \caption{Analytical bias for ({\it left}) a two beam window function with $W_l^S(20,30)$ and ({\it right}) a Gaussian window function with  $W_l^G(40)$ computed from the average $C_l$ from  1000 realizations of a SI CMB map compared with $\langle \kappa_l^{realization}\rangle$ (the average $\kappa_l$ from 1000 realizations). This shows that the theoretical bias is a very good estimation of the bias for a statistically isotropic map. }
  \label{bias3020}
\end{figure}

In statistics, an estimator is a function of the known data that is used to estimate an unknown parameter; an estimate is the result from the actual application of the function to a particular set of data. Many different estimators may be possible for any given parameter. 
The above theory can be use to construct an  estimator for measuring BipoSH coefficients from a given CMB map
\be \label{estimator}
\tilde A_{ll^\prime}^{\ell M} = \sum_{m m^\prime} \sqrt{W_l W_{l'}} a_{lm}a_{l^\prime
m^\prime} \, \, {\mathcal{ C}}^{\ell M}_{lml^\prime m^\prime}\,\quad ,
\ee
where $W_l$ is the Legendre transform of the window function. The above estimator 
is un-biased.  
Bias is the mismatch between ensemble average of the estimator and the true value. Bias for the BiPS is defined as ${\mathfrak B}_\ell=\langle\tilde\kappa_\ell\rangle-\kappa_\ell$ and is equal to
\begin{eqnarray}\label{klbias}
{\mathfrak B}_\ell = \sum_{l_1,l_2}W_{l_1}
W_{l_2}\,\, \sum_{m_1,m_1^\prime} \sum_{m_2,m_2^\prime}&&\,\left[\langle
a^*_{l_1m_1}a_{l_1 m_1^\prime}\rangle\langle a^*_{l_2m_2}a_{l_2
m_2^\prime}\rangle + \langle a^*_{l_1m_1}a_{l_2
m_2^\prime}\rangle\langle a^*_{l_2m_2}a_{l_1 m_1^\prime}\rangle
\right] \nonumber \\ &&{}\times \sum_M \mathcal{ C}^{\ell
M}_{l_1m_1l_2m_2}\mathcal{ C}^{\ell M}_{l_1m_1^\prime
l_2m_2^\prime}\,.
\end{eqnarray}
Therefore an un-biased estimator of BiPS is given by
\be
\tilde\kappa_\ell = \sum_{ll^\prime M} \left|\tilde A_{ll^\prime}^{\ell
M}\right|^2 - {\mathfrak B}_\ell\, ,
\ee
The above expression for ${\mathfrak B}_\ell$ is obtained by assuming Gaussian statistics of the temperature fluctuations. The procedure is very similar to computing cosmic variance (which is discussed in the next section), but much simpler. However, we can not measure the ensemble average in the above expression and as a result, elements of the covariance matrix (obtained from a single map) are poorly determined due to the cosmic variance. The best we can do is to compute the bias for the SI component of a map 
\begin{equation}\label{klisobias}
{\mathfrak B}_\ell \equiv\langle\tilde\kappa_\ell^B\rangle_{_{\rm SI}}
 = (2\ell+1)\,\sum_{l_1} \sum_{l_2=|\ell-l_1|}^{\ell+l_1} C_{l_1}
 C_{l_2} W_{l_1} W_{l_2} \left[1 + (-1)^{\ell}\, \delta_{l_1
 l_2}\right]\,.
\end{equation}
{\em Note , the estimator $\tilde \kappa_\ell$ is unbiased, only for
SI correlation,i.e., $\langle \tilde \kappa_\ell \rangle=0$.}
Consequently, for SI correlation, the measured $\tilde \kappa_\ell$
will be consistent with zero within the error bars given by
$\sigma_{_{\rm SI}}$ \cite{us_apjl}.  We simulated 1000 SI CMB maps and computed BiPS for them using different filters. The average BiPS of SI maps is an estimation of the bias which can be compared to our analytical estimation. Fig. \ref{bias3020} shows that the theoretical bias (computed from average $C_l$) match the numerical estimations of average $\kappa_{\ell}$ of the 1000 realizations of the SI maps.

It is important to note that bias cannot be correctly subtracted for
non-SI maps. Non-zero $\tilde \kappa_\ell$ estimated from a non-SI map
will have contribution from the non-SI terms in full bias given in
eq.~(\ref{klbias}). It is not inconceivable that for strong SI
violation, ${\mathfrak B}_\ell$ over-corrects for the bias leading to
negative values of $\tilde \kappa_\ell$. {\em What is important is
whether measured $\tilde \kappa_\ell$ differs from zero at a
statistically significant level.}

\section{Cosmic Variance}
Cosmic variance is defined as the variance of the estimator of the BiPS 
\be
\sigma^2\,=\,<\tilde{\kappa}_{\ell}^2>-<\tilde{\kappa}_{\ell}>^2
\ee
We can analytically compute the variance of $\tilde{\kappa}^{\ell}$ using the Gaussianity of $\Delta T$. Looking back at the eq.(\ref{kl}) we can see, we will have to calculate the eighth moment of the field 
\be
\langle \Delta T(\hat{n}_1)\Delta T(\hat{n}_2)\Delta T(\hat{n}_3)\Delta T(\hat{n}_4)\Delta
T(\hat{n}_5)\Delta T(\hat{n}_6)\Delta T(\hat{n}_7)\Delta T(\hat{n}_8) \rangle.
\ee
Assuming Gaussianity of the field we can rewrite the 
eight point correlation in terms of two point correlations. 
One can write a simple code to do that{\footnote{F90 software implementing this is available from the authors upon request}}. This  will give us $(8-1)!!=7\times 5 \times 3=105$ terms. These 105 terms consist of terms like:
\be
\langle \Delta T(\hat{n}_1)\Delta T(\hat{n}_2) \rangle\langle\Delta T(\hat{n}_3)\Delta T(\hat{n}_4) \rangle\langle\Delta T(\hat{n}_5)\Delta T(\hat{n}_6) \rangle\langle\Delta T(\hat{n}_7)\Delta T(\hat{n}_8) \rangle, 
\ee
and all other permutations of them. 
On the other  hand $\langle\tilde{\kappa}_{\ell}\rangle$ has  a 4 point correlation in it which can also be expanded versus two point correlation functions. 
If we form $\langle \tilde{\kappa}_{\ell}^2\rangle-\langle\tilde{\kappa}_{\ell}\rangle^2$, only 96 terms will be left which are in the following form
\be
\label{firstterm}
(\frac{2\ell+1}{8\pi^2})^2 \int d\Omega_1 \cdots d\Omega_4 
\int dR \int dR' \chi_{\ell}(R) \chi_{\ell}(R')  C(\hat{n}_1,R'\hat{n}_4)C(\hat{n}_2,R'\hat{n}_3)C(R\hat{n}_1,\hat{n}_4)C(R\hat{n}_2,\hat{n}_3)
\ee
and all other permutations. In Appendix \ref{AppCosVar} we explain how we can handle these 96 terms to compute the following analytical expression for cosmic variance
\begin{eqnarray}\label{klcv}
&&\sigma^2_{_{\rm SI}}(\tilde\kappa_\ell) =\sum_{l : 2l \ge \ell}\!\!
4\, C_{l}^4 W_l^4 \left[ 2 \frac{(2\ell+1)^2}{2l+1}+ (-1)^{\ell}
(2\ell+1)+ (1+2(-1)^{\ell}) F_{ll}^\ell\right] \nonumber \\
&&{}+\sum_{l_1} \sum_{l_2=|\ell-l_1|}^{\ell+l_1} \!\!\! 4
\,C_{l_1}^2\,C_{l_2}^2 W_{l_1}^2\,W_{l_2}^2\left[ (2\ell+1) +
F_{l_1l_2}^\ell \right]\nonumber \\
&& + \,8\sum_{l_1}\,\frac{(2\ell+1)^2}{2l_1+1}
\,C_{l_1}^2 W_{l_1}^2 \left[\sum_{l_2=|\ell-l_1|}^{\ell+l_1} C_{l_2}
W_{l_2}\right]^2\nonumber \\
&&{} + 16\,(-1)^{\ell}\,\sum_{l_1 : 2l_1 \ge \ell}
\,\frac{(2\ell+1)^2}{2l_1+1}\, \sum_{l_2=|\ell-l_1|}^{\ell+l_1}
C_{l_1}^3 C_{l_2}\,W_{l_1}^3 W_{l_2}.
\end{eqnarray}

\begin{figure}[p]
  \includegraphics[scale=0.3, angle=-90]{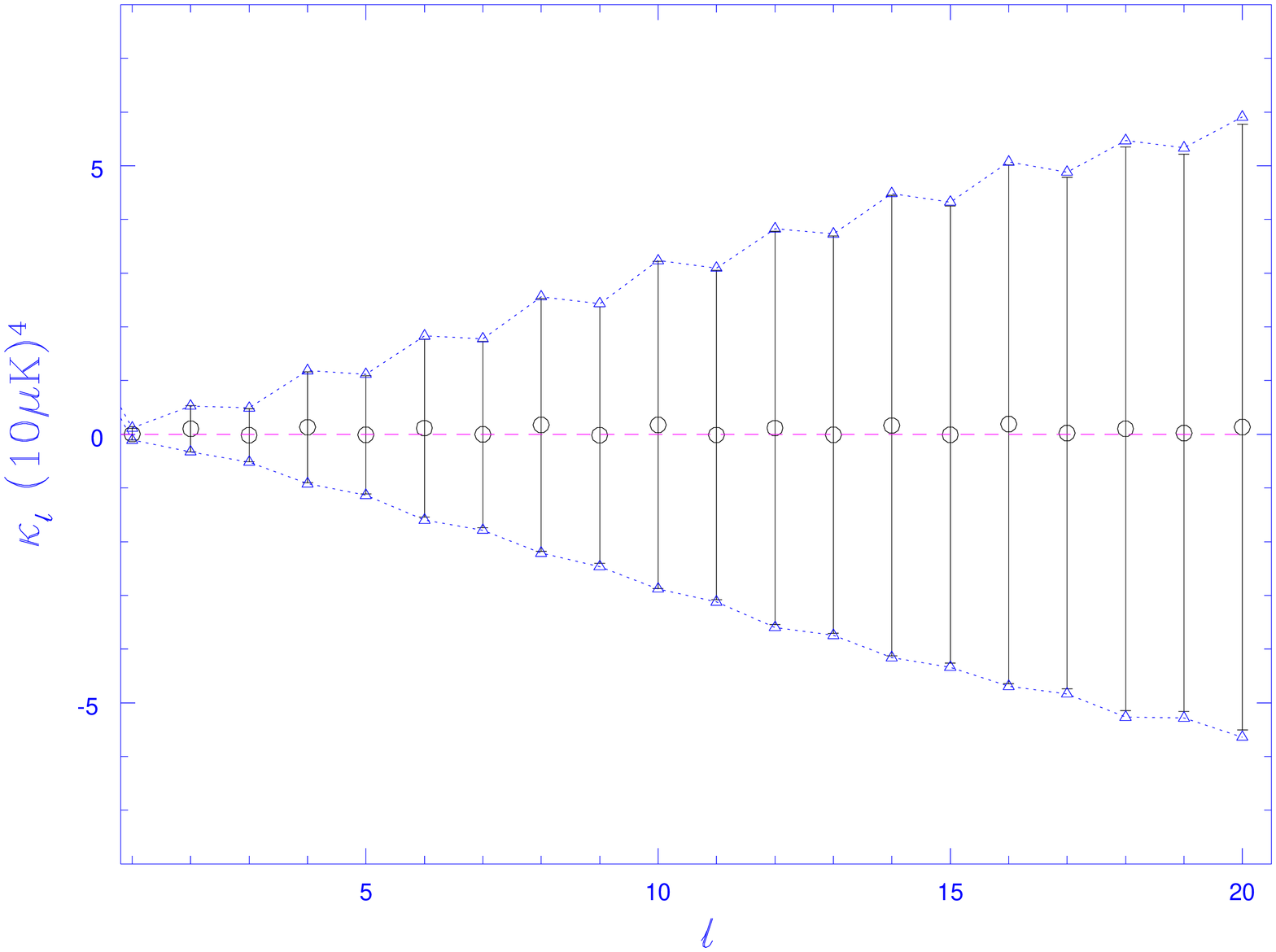}
  \includegraphics[scale=0.3, angle=-90]{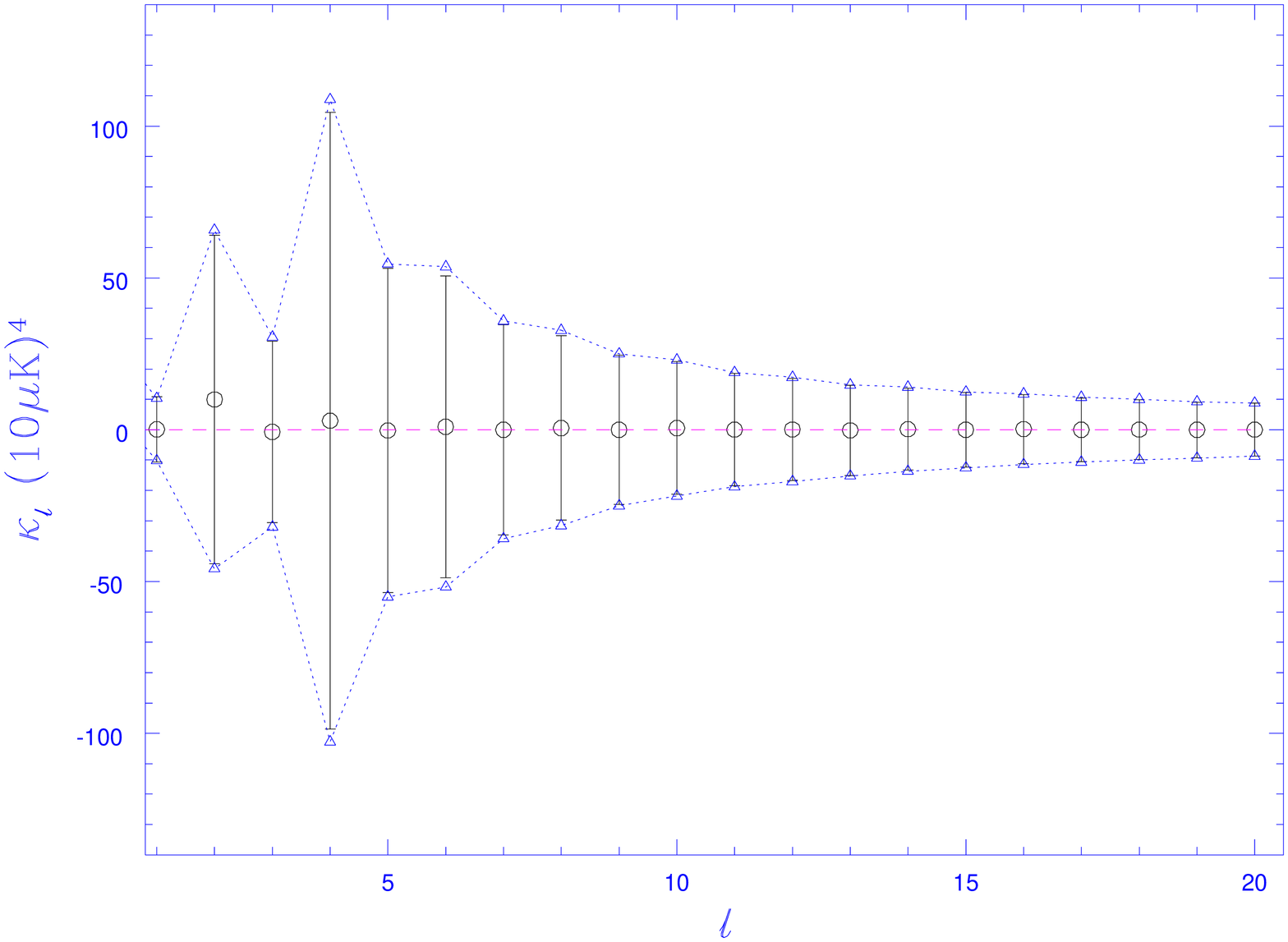}
  \caption{ Bias corrected `measurement' of BiPS,
 $\tilde{\kappa_\ell}$ for SI CMB maps with a best fit LCDM power spectrum
 smoothed by ({\em left:}) a two beam function $W_l^S(20,30)$ and ({\em right:}) a  Gaussian beam $W_l^G(40)$. The
 cosmic error, $\sigma({\kappa_\ell})$, obtained using $1000$
 independent realizations of CMB (full) sky map matches the analytical
 results shown by dotted curve with triangles . This shows a much better fit
 to the theoretical cosmic variance compared to what was obtained for $100$
 realizations \cite{us_apjl}}
  \label{comparison3020}
  \label{comparison400}
\end{figure}

Numerical computation of $\sigma^2_{_{\rm SI}}$ is fast. But the challenge is to compute Clebsch-Gordan coefficients for large quantum numbers. We use drc3j
subroutine of netlib\footnote{http://www.netlib.org/slatec/src/} in order to compute the Clebsch-Gordan coefficients in our codes. Again we can check the accuracy of our analytical estimation of cosmic variance by comparing it against the standard deviation of BiPS of 1000 simulations of SI CMB sky. The result is shown in Fig. \ref{comparison3020} and shows a very good agreement between the two.

\section{Sources of Breakdown of Statistical Isotropy}
An observed map of CMB anisotropy, $\Delta T^{obs}(\hat{n})$, is an $n$-dimensional vector ($n=12\,\times\,512^2$ for the WMAP data at HEALPix resolution $N_{\mathrm side}=512$).  This observed map contains the true CMB temperature fluctuations, $\Delta T(\hat{n})$, convolved with the beam and buried into noise and foreground contaminations. The observed map is related to the true map through this relation
\be
{\mathbf \Delta T^{obs}}\,=\,B {\mathbf \Delta T}\,+\,\mathbf{n},
\ee
in which $B$ is a matrix that contains the information about the beam smoothing effect and $\mathbf{n}$ is the contribution from instrumental noise and foreground contamination. Hence, the observed map is a realization of a Gaussian process with covariance $C=C^T+C^N+C^{res}$ where $C^T$ is the theoretical covariance of the CMB temperature fluctuations, $C^N$ is the noise covariance matrix and $C^{res}$ is the covariance of residuals of foregrounds.
Breakdown of statistical isotropy can occur in any of these parts.  In general we can divide these effects into two kinds:
\begin{itemize}
\item{Theoretical effects:} These effects are theoretically motivated and are intrinsic to the true CMB sky,  ${\Delta T}$. We discuss two examples of these effects, {\it i.e.} non-trivial cosmic topology and primordial magnetic fields, in the next sections. 
\item{Observational artifacts:} In an ideally cleaned CMB map, the true CMB temperature fluctuations are completely extracted from the observed map. But this is not always true. Sometimes there are some artifacts (related to $B$ or $\mathbf{n}$) left in the cleaned map which may in principle violate the SI. These effects are explained in section \ref{artifacts}.
\end{itemize}
\subsection{Signatures of Cosmic Topology}
The question of size and the shape of our universe are very old problems that have received increasing attention over the past few years \cite{ell71, lac_lum95, levin98, bps, angelwmap, staro, coles-graca, cop04}. Although a multiply connected universe sounds non-trivial, but there are theoretical motivations \cite{linde, lev02} to favor a spatially compact universe. One possibility to have a compact flat universe is the consideration of multiply connected (topologically nontrivial) spaces. 
The oldest way of searching for global structure of the universe is by identifying ghost images of local galaxies and clusters or quasars at higher redshifts \cite{lac_lum95}. This method can  probe the topology of the universe only on scales substantially smaller than the apparent radius of the observable universe.  Another method to search for the shape of the universe is through the effect on the power spectrum of cosmic density perturbation fields. In principle, this effect can be observed in the distribution of matter in the universe and the CMB anisotropy. 
Over the past few years, many independent methods have been proposed to search  for evidence of a finite universe in CMB maps. These methods can be classified in three main groups.
\begin{itemize}
\item Using the angular power spectrum of CMB anisotropies to probe the topology of the Universe. The angular power spectrum, however, is inadequate to characterize the peculiar form of the anisotropy manifest in small universes of this type. It has been shown that a nontrivial topology breaks the SI down and as a result, there is more information in a map of temperature fluctuations than just the angular power spectrum \cite{levin98, bps, us_prl}.
\item The second class of methods are direct methods which rely on multiple imaging of the CMB sky.  The most well known methods amongst these methods are S-map statistics \cite{staro, angelwmap} and the search for circles-in-the-sky \cite{circles}. 
\item Third class of methods are indirect probes which deal with the  correlation patterns of the CMB anisotropy field by using an appropriate combination of coefficients of the harmonic expansion of the field \cite{coles-graca, donoghue, us_prl, cop04}.
\end{itemize}
 The BiPS method is one of the strategies for detecting observable signatures of a small compact universe. 
The BiPS method is essentially based on the search for correlation patterns in CMB.   Using the fact that statistical isotropy is violated in compact spaces one could use the  bipolar power spectrum as a probe to detect the topology of the universe. A simple example of application of BiPS in constraining topology of the universe is for a $T^3$ universe, where the correlation function is given by 
 \begin{equation}
C({\hat q,\hat q^\prime}) = L^{-3} \sum _{{\bf n}}
P_\Phi(k_{\bf n}) \,\,{\mathrm e}^{-i \pi 
(\epsilon_{\hat q} {\bf n}\cdot {\hat q} - \epsilon_{\hat q^\prime} {\bf n}\cdot 
{\hat q}^\prime)},
\label{C_tor}
\end{equation}
in which, ${\bf n}$ is 3-tuple of integers (in order to avoid confusion, we use $\hat{q}$ to represent the direction instead of $\hat{n}$),  the small parameter $\epsilon_{\hat q} \le 1 $ is the physical
distance to the SLS along $\hat q$ in units of $L/2$ (more generally,
$\bar L/2$ where $\bar L= (L_1L_2L_3)^{1/3}$) and $L$ is the size of the Dirichlet domain (DD). When $\epsilon$ is a small
constant, the leading order terms in the correlation function
eq.~(\ref{C_tor}) can be readily obtained in power series expansion in
powers of $\epsilon$.  For the lowest wavenumbers $|{\mathbf n}|^2=1$
in a cuboid torus
\bea
C({\hat q,\hat q^\prime}) &\approx& 2 \sum_i P_\Phi({2\pi}/{L_i}) 
\cos(\pi\epsilon\beta_i\Delta q_i) \\ \nonumber
&\approx& C_0 \left[1 - \epsilon^2\, 
|{\mathbf \Delta q}|^2 + 3\,\epsilon^4 \, \sum_{i=1}^3
(\Delta q_i)^4   \right],
\label{appcorr}
\eea
where $\Delta q_i$ are the components of ${\mathbf \Delta q} =\hat
q-\hat q^\prime$ along the three axes of the torus and $\beta_i = \bar
L /L_i$. From this, the non-zero $\kappa_\ell$ can be analytically computed
to be
\begin{eqnarray}
\frac{\kappa_0}{C_0^2}\, &=&\,\pi^2(1-4\epsilon^2 
+\frac{368}{15} \epsilon^4-\frac{288}{5}\epsilon^6+\frac{20736}{125}\epsilon^8)
\nonumber \\
\frac{\kappa_4}{C_0^2}\, &=&\, \frac{12288 \pi^2 }{875} \epsilon^8  
\end{eqnarray}
$\kappa_4$ has the information of the relative size of the Dirichlet domain and one can use it to constrain the topology of the universe. A detailed description of determining the topology  of the universe with BiPS is given in \cite{us_prl}. 

When CMB anisotropy is multiply imaged, the bipolar power  spectrum corresponds to a correlation pattern of matched pairs of circles.    Since it is orientation independent, the BiPS has to be computed only once for the  CMB map irrespective of the orientation of the DD with respect to the sky. This method is very fast because there are fast spherical harmonic transform methods  of the CMB sky maps. Using these methods it is very fast and efficient to compute the BiPS even for maps in WMAP resolution (HEALPix, $N_{side}=512$) in a few minutes on a single processor. We have used BiPS to put constraints on the topology of the Universe. The results are in preparation and will be reported soon\cite{us_dodeca}.

 The $A^{\ell M}_{l_1l_2}$ signature, which is not discussed here, contains more details of orientation of the SI violation. It may be possible to detect  the $A^{\ell M}_{l_1l_2}$ too for strong violation of SI.

\subsection{Primordial Magnetic Fields}
It has been shown that a cosmological magnetic field, generated during an early epoch of inflation~\cite{ratra1992, bamba2004},  can generate CMB anisotropies~\cite{DKY}. The presence of a preferred direction due to a homogeneous magnetic field background leads to non-zero off-diagonal elements in the covariance matrix~\cite{gang}. This induces correlations between $a_{l+1,m}$ and $a_{l-1,m}$  multipole coefficients of the CMB temperature anisotropy field in the following manner
 \be \label{alfven}
 \langle a_{lm}a^*_{l^\prime m^\prime}\rangle\,=\,
 \delta_{m,m'} [\delta_{l,l'}C_l+(\delta_{l+1,l'-1}+\delta_{l-1,l'+1}D_l)],
 \ee
 where $D_l$ is the power spectrum of off-diagonal elements of the covariance matrix. For a Harrison-Peebles-Yu-Zel'dovich scale-invariant spectrum, $D_l$ behaves as $l^{-2}$. More precisely, it is given by
 \be
 D_l\,=\,4\times 10^{-16} l^{-2} (\frac{B}{1nG})^4.
 \ee
 This clearly violates the statistical isotropy.  If we substitute the above $\langle a_{lm}a^*_{l^\prime m^\prime}\rangle$ into eqn.~(\ref{ALMvsalm}), we will get
 \bea
 A_{l_1l_2}^{{\ell}M} &\,=\,& \delta_{M,0} [(-1)^{l_1}  \sqrt(2l_1+1)  C_{l_1} \delta_{l_1,l_2}  \delta_{\ell,0} \\ \nonumber
 &\,+\,&
 D_{l_1} \sum_{m1=-l1,l1}(-1)^{m_1}C_{l_1m_1l_2-m_1}^{{\ell}M}\delta_{l_{1+1},l_{2-1}}
\\ \nonumber
&\,+\,&
D_{l_1} \sum_{m1=-l1,l1}(-1)^{m_1}C_{l_1m_1l_2-m_1}^{{\ell}M}\delta_{l_{1-1},l_{2+1}} ].
\eea

The first line is the statistically isotropic part and it just contributes to $\kappa_0$. But the interesting parts are the two other lines. We can substitute this expression into eqn. (\ref{kappal}) and compute the BiPS predictions for magnetic fields. This work is still under process and will be reported in the near future \cite{us_magneticfield}.

\subsection{Observational Artifacts} \label{artifacts}
Foregrounds and observational artifacts (such as non-circular beam, incomplete/non-uniform sky coverage and anisotropic noise) would also manifest themselves as violations of SI. 
\begin{itemize}
\item {\sf Anisotropic noise~:} The CMB temperature measured by an
  instrument is a linear sum of the cosmological signal as well as
  instrumental noise. The two point correlation function then has two
  parts, one part comes from the signal and the other one comes from
  the noise 

\be C(\hat{n}_1,\, \hat{n}_2)\, =\, C^S(\hat{n}_1,\,
  \hat{n}_2)\,+\,C^N(\hat{n}_1,\, \hat{n}_2).  
\ee 

Both signal and noise should be statistically isotropic to have a
statistically isotropic CMB map.  So even for a statistically
isotropic signal, if the noise fails to be statistically isotropic the
resultant map will turn out to be anisotropic. The noise matrix can
fail to be statistically isotropic due to non-uniform coverage. Also
if the noise is correlated between different pixels the noise matrix
could be statistically anisotropic. A simple example of this is the diagonal
 (but anisotropic) noise given by the following correlation
\be
C^N(\hat{n},\hat{n}')=\sigma^2(\hat{n}) \delta_{\hat{n}\hat{n}'}.
\ee
This noise clearly violates the SI and will result  a non-zero BiPS given by
\be
\kappa_{\ell}=\sum_{m=-\ell}^{\ell}{|f_{\ell m}|^2},
\ee
where $f_{\ell m}$ are spherical harmonic transform of the noise, $f_{\ell m}=\int{\mathrm{d} \Omega_{\hat{n}}Y_{\ell m}^{*}(\hat{n})\sigma^2(\hat{n})}$. 

\item {\sf The effect of non-circular beam~:} In practice when we deal
  with  data, it is necessary to take into account the {\it
    instrumental response}. The instrumental response is nothing but
  the beams width and the form of the beam and can be taken into
  account by defining a beam profile function $B(\hat{n},\,
  \hat{n}')$. Here $\hat{n}$ denotes the direction to the center of
  the beam and $\hat{n}' $ denotes the direction of the incoming
  photon. The temperature measured by the instrument is given by \be
  \Delta \tilde{T}(\hat{n})\,=\,\int \Delta T (\hat{n}')B(\hat{n},\,
  \hat{n}')d\Omega_{\hat{n}'} \ee Using this relation to calculate the
  correlation function $\tilde{C}(\hat{n}_1,\, \hat{n}_2)\, =\,
  \langle \Delta \tilde{T}(\hat{n}_1) \Delta
  \tilde{T}(\hat{n}_2)\rangle$ one would get \bea
  \tilde{C}(\hat{n}_1,\, \hat{n}_2)\,&=&\,\int d\Omega_{\hat{n}'} \int
  d\Omega_{\hat{n}''} \langle \Delta T(\hat{n}') \Delta
  T(\hat{n}'')\rangle B(\hat{n}_1,\, \hat{n}')B(\hat{n}_2,\,
  \hat{n}'') \\ \nonumber &=&\,\int d\Omega_{\hat{n}'} \int
  d\Omega_{\hat{n}''} C(\hat{n}',\hat{n}'') B(\hat{n}_1,\,
  \hat{n}')B(\hat{n}_2,\, \hat{n}'').  \eea 
  Only for a circular beam
  where $B(\hat{n},\, \hat{n}')\, \equiv \,B(\hat{n}\cdot \hat{n}')$,
  the correlation function is statistically isotropic,
  $\tilde{C}(\hat{n}_1,\, \hat{n}_2)\,\equiv\,\tilde{C}(\hat{n}_1
  \cdot \hat{n}_2)$.  Breakdown of SI is obvious since even $C_l$ get 
  mixed  for a non-circular beam, $\tilde{C}_l=\sum_{l'}{A_{ll'}C_{l'}}$ 
  \cite{beam}. Non-circularity of the beam in CMB anisotropy
  experiments is becoming increasingly important as experiments go for
  higher resolution measurements at higher sensitivity. 
\item{\sf Mask effects~:} Many experiments  map only a part of the sky. 
  Even in the best case,  contamination by galactic  foreground residuals 
  make parts of the sky unusable. 
  The incomplete sky or mask effect is
  another source of breakdown of SI.  But, this effect can be readily
  modeled out. The effect of a general mask on the temperature field is as follows
\be
\Delta T^{masked}(\hat{n})\, =\, \Delta T(\hat{n}) W(\hat{n}),
\ee
where $W(\hat{n})$ is the mask function. One can cut different parts of the sky by choosing appropriate mask functions. Masked $a_{lm}$ coefficients can be computed from the masked temperature field,
\bea \label{maskedalm}
a_{lm}^{masked}
\,&=&\,\int{\Delta T^{masked}(\hat{n}) Y_{lm}^*(\hat{n}) d\Omega_{\hat{n}}}
\\ \nonumber
&=&\, \sum_{l_1 m_1}{a_{l_1 m_1} \int{Y_{l_1 m_1}(\hat{n}) Y_{lm}^*(\hat{n}) W(\hat{n}) d\Omega_{\hat{n}}}}.
\eea
Where $a_{l_1 m_1}$ are spherical harmonic transforms of the original temperature field. We can expand $W(\hat{n})$ in spherical harmonics as well
\be 
W(\hat{n})\,=\,\sum_{l m}{w_{lm} Y_{l m}(\hat{n})},
\ee 
and after substituting this into eqn.~(\ref{maskedalm}) it is seen that  the
masked $a_{lm}$ is given by the effect of a kernel $K_{lm}^{l_1 m_1}$ on original $a_{lm}$ \cite{prunet}
\be
a_{lm}^{masked}\,=\, \sum_{l_1 m_1}{a_{l_1 m_1} K_{lm}^{l_1 m_1}}.
\ee
The kernel contains the information of our mask function and is defined by
\bea
K_{lm}^{l_1 m_1}\,&=&\,\sum_{l_2m_2}{w_{l_2m_2}\int{Y_{l_1 m_1}(\hat{n}) Y_{l_2 m_2}(\hat{n})Y_{lm}^*(\hat{n}) d\Omega_{\hat{n}}}}
\\ \nonumber
&=&\,\sum_{l_2m_2}{w_{l_2m_2}\sqrt{\frac{(2l_1+1)(2l_2+1)}{4\pi(2l+1)}}C_{l_10l_20}^{l0}C_{l_1m_1l_2m_2}^{lm}}.
\eea
In the last step of the above expression we used rule \ref{A6} of spherical harmonics to simplify the expression. The covariance matrix of a masked sky will no longer have the diagonal form of eqn.(\ref{diagonalcovmat}) because of the action of the kernel
\bea
\langle a^{masked}_{lm}a^{masked\, *}_{l^\prime m^\prime}\rangle
\,&=&\,\langle a_{l_1m_1}a^{*}_{l^\prime_1 m^\prime_1}\rangle 
K_{lm}^{l_1 m_1}K_{l^\prime m^\prime}^{l^\prime_1 m^\prime_1}\\ \nonumber
\,&=&\,C_{l_1}\delta_{l_1l^\prime_1}\delta_{m_1m^\prime_1}K_{lm}^{l_1 m_1}K_{l^\prime m^\prime}^{l^\prime_1 m^\prime_1}\\ \nonumber
\,&=&\,\sum_{l_1,m_1}{C_{l_1}K_{lm}^{l_1 m_1} K_{l^\prime m^\prime}^{l_1 m_1}}.
 \eea
This clearly violates the SI and results a non-zero BiPS for masked CMB skies. 
 In the next section we apply a galactic mask to ILC map and  show that signature of this mask on BiPS is a rising tail at low $\ell, (\ell < 20)$. 
\item{\sf Residuals from foreground removal~:} Besides the
  cosmological signal and instrumental noise, a CMB map also contains
  foreground emission such as galactic emission, etc. The foreground
  is usually modeled out using spectral information. However,
  residuals from foreground subtractions in the CMB map will violate SI. Interestingly, BiPS does sense the difference between maps with grossly different emphasis on the galactic foreground. As an example we compute BiPS of a Wiener filtered map in the next section to make this point clear (see Fig. \ref{wiener}). It shows a signal very similar to that of a galactic cut sky. This can be understood if one writes the effect of the Wiener filter as a weight on some regions of the map
\be 
\Delta T^{W}(\hat{n})\, =\, \Delta T(\hat{n}) (1+W(\hat{n})).
\ee
This explains the similarity between a cut sky and a Wiener filtered map. 
The effect of foregrounds on BiPS still needs to be studied more carefully.
\end{itemize}

\section{BiPS Results from WMAP} 
\begin{figure*}[h]
  \includegraphics[scale=0.6, angle=-90]{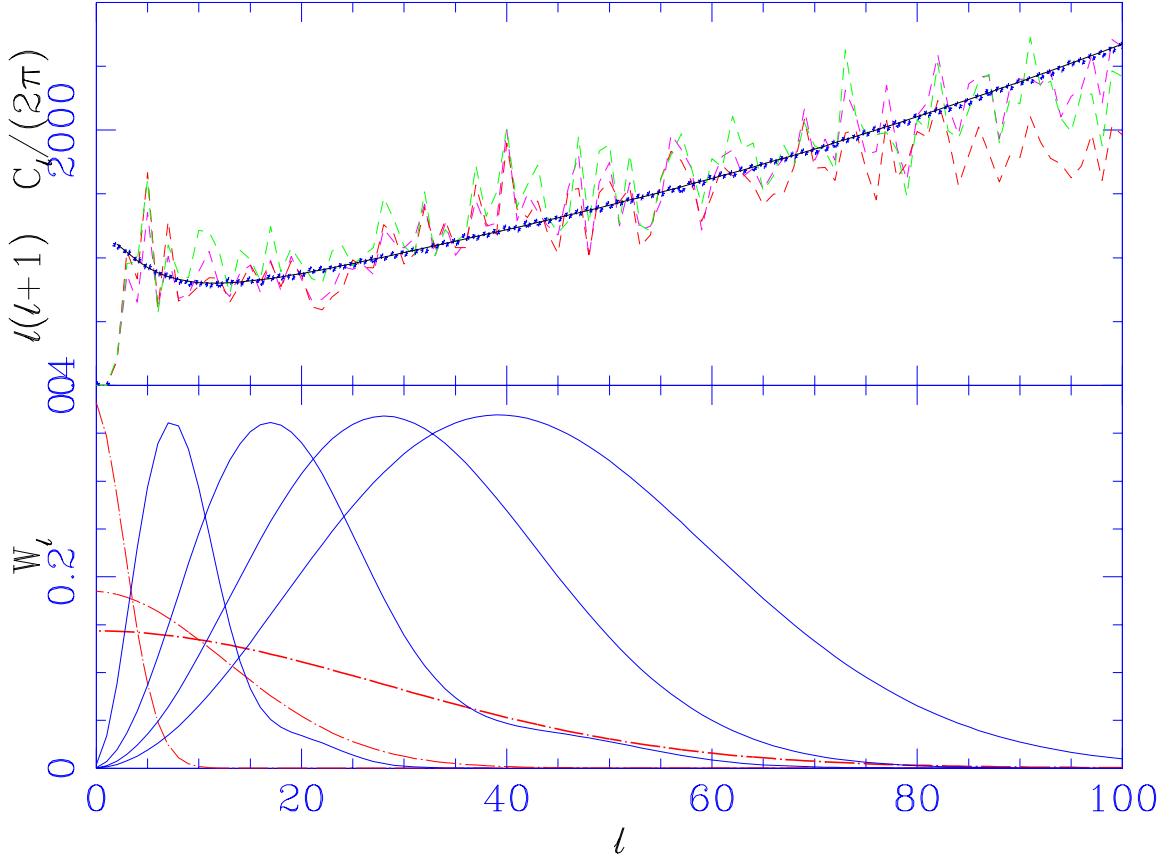}
 \caption{  {\em Top:} 
$C_{\ell}$ of the two WMAP CMB anisotropy maps. The red, magenta and green curves correspond to map A, B and C, respectively. The black line is a `best fit' WMAP theoretical $C_{\ell}$ used for simulating SI maps. Blue dots are the average $C_l$ recovered from $1000$ realizations. {\em Bottom:} These plots show
 the window functions used. The dashed curves with increasing $l$ coverage are `low-pass' filter, $W_l^G(l_s)$, with $l_s=4, 18, 40$, respectively.  The solid lines are `band-pass' filter $W^S_l(l_t,l_s)$ with $(l_s,l_t)=(13,2), (30,5), (30,20), (45,20)$, respectively.}
\label{cl_wmap_cl100}
\end{figure*}

We carry out measurement of the BiPS, on the following 
CMB anisotropy maps
\begin{itemize}
\item[A)] a foreground cleaned map (denoted as `TOH')~\cite{maxwmap},
\item[B)] the Internal Linear Combination map (denoted as `ILC' in the figures)~\cite{wmap}, and
\item[C)] a customized linear combination of the QVW maps of WMAP with a galactic cut (denoted as `CSSK').
\end{itemize}
Also for comparison, we measure the BiPS of 
\begin{itemize}
\item[D)] a Wiener filtered map of WMAP data (denoted as `Wiener')~\cite{maxwmap}, and
\item[E)] the ILC map with a $10^{\circ}$ cut around the equator (denoted as `Gal. cut.').
\end{itemize}
Angular power spectra of these maps are shown in Fig.~\ref{cl_wmap_cl100}. The best fit theoretical power spectrum from the WMAP analysis~\footnote{Based on an LCDM model with a scale-dependent (running) spectral index which
best fits the dataset comprised of WMAP, CBI and ACBAR CMB data
combined with 2dF and Ly-$\alpha$ data} ~\cite{sper_wmap03} is plotted on the same figure. 
$C_l$ from observed maps are consistent with the theoretical curve, $C_l^T$, (except for low multipoles. The bias and cosmic variance of BiPS depend on the total SI angular power
spectrum of the signal and noise $C_l = C_l^S + C_l^N$.  However, we have
restricted our analysis to $l \lsim 60$ where the errors in the WMAP power spectrum is
dominated by cosmic variance.  It is conceivable that the SI violation is
limited to particular range of angular scales.  Hence, multipole space
windows that weigh down the contribution from the SI region of
multipole space will enhance the signal relative to cosmic
error, $\sigma_{_{\rm SI}}$. We use simple filter functions in $l$ space to isolate different ranges of angular scales; a low pass, Gaussian filter 
\begin{equation}
W^G_l(l_s) = \exp(-(l+1/2)^2/(l_s+1/2)^2)
\label{gaussfilter}
\end{equation}
that cuts off power on small angular scales ($\lsim 1/l_s$) and a band
pass filter,
\begin{equation}
W^S_l(l_t, l_s) = \left[2(1- J_0((l+1/2)/(l_t+1/2)))
\right]\,\exp(-(l+1/2)^2/(l_s+1/2)^2)
\label{bpfilter}
\end{equation}
that retains power within a range of multipoles set by $l_t$ and
$l_s$.  The windows are normalized such that $\sum_l (l+1/2)/(l(l+1))
W_l =1$, i.e., unit {\it rms} for unit flat band power
$C_l=1/(l(l+1))$. The window functions used in our work are plotted in
figure~\ref{cl_wmap_cl100}.
We use the  $C_l^T$ to generate 1000 simulations of the SI CMB
maps. $a_{lm}$'s are generated up to an $l_{max}$ of 1024
(corresponding to HEALPix resolution $N_{side}=512$). These are then
multiplied by the window functions $W^G_l(l_s)$ and $W^S_l(l_t,
l_s)$. We compute the BiPS for each realization.  Fig.\ref{cl_wmap_cl100} shows that the average power spectrum obtained from the simulation matches
the theoretical power spectrum, $C_l^T$, used to generate the
realizations. We use $C_l^T$ to analytically compute bias and 
cosmic variance estimation for $\tilde{\kappa_{\ell}}$. This allows us to rapidly compute BiPS with $1\sigma$ error bars for different theoretical $C_l^T$.

\begin{figure}[p]
  \includegraphics[scale=0.6, angle=-90]{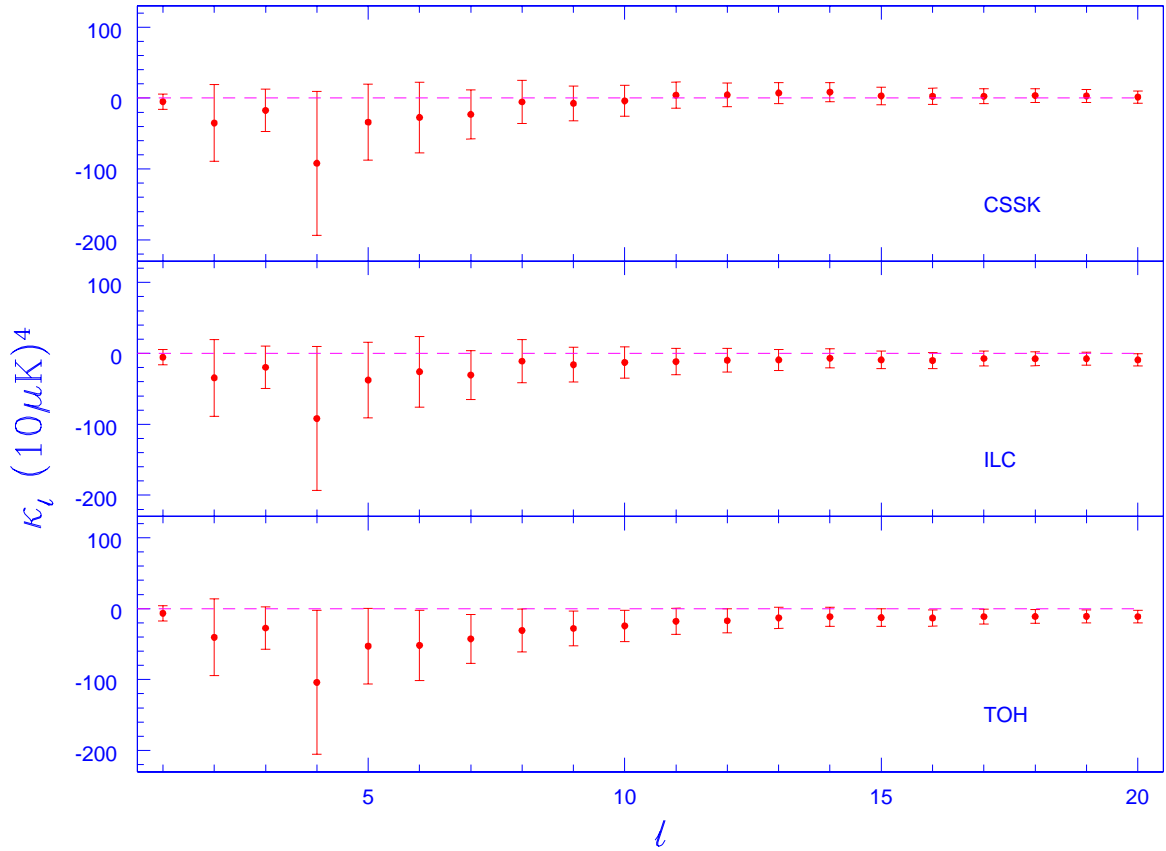}
  \caption{Figure shows the measured values of
$\kappa_\ell$ for maps A, B and C filtered with a Gaussian window with
 $l_s =40$. Each map appears to show the violation of statistical isotropy for some $l$ at the level of $\sim 1 \sigma$. However, this is because the PDF for these models are skewed with the maximum probability shifted away from the mean to negative values (see Fig. \ref{pdf1to5}).}
  \label{kappa_wmap_400}
\end{figure}

\begin{figure}[h]
  \includegraphics[scale=0.7, angle=-90]{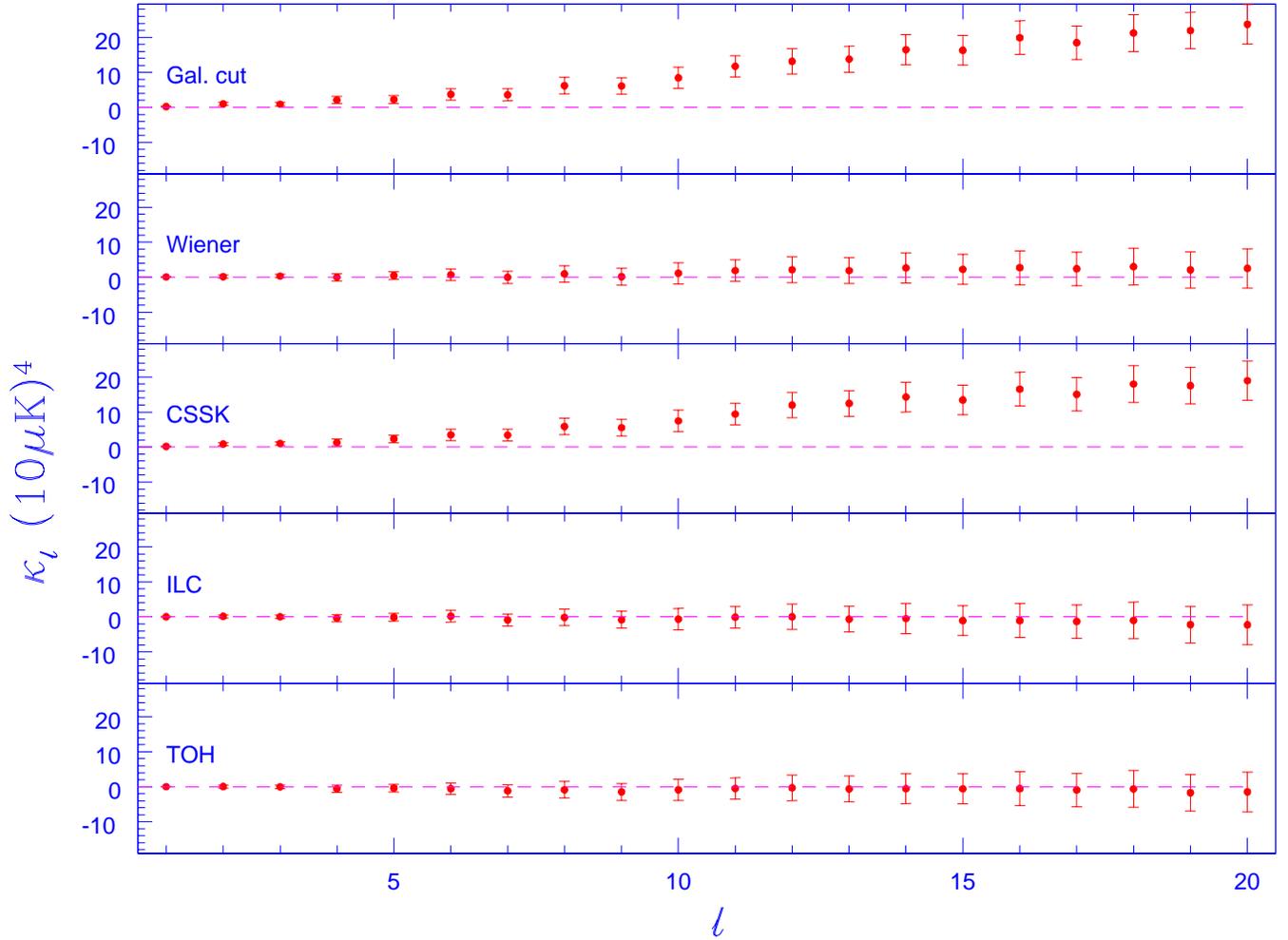}
  \caption{Measured BiPS for maps A, B and C filtered with a window with
$l_s =30$, $l_t =20$. This is to check the  statistical isotropy of the WMAP in the modest $20<l<40$ in the multipole space where certain anomalies have been reported. ILC with a 10-degree-cut (top) has the same BiPS as map C ($l_s =30$, $l_t =20$) which explains that the raising tail of CSSK map is because of the mask. }
  \label{kappa_wmap_3020}
\end{figure}

\begin{figure}[h]
   \includegraphics[scale=0.6, angle=-90]{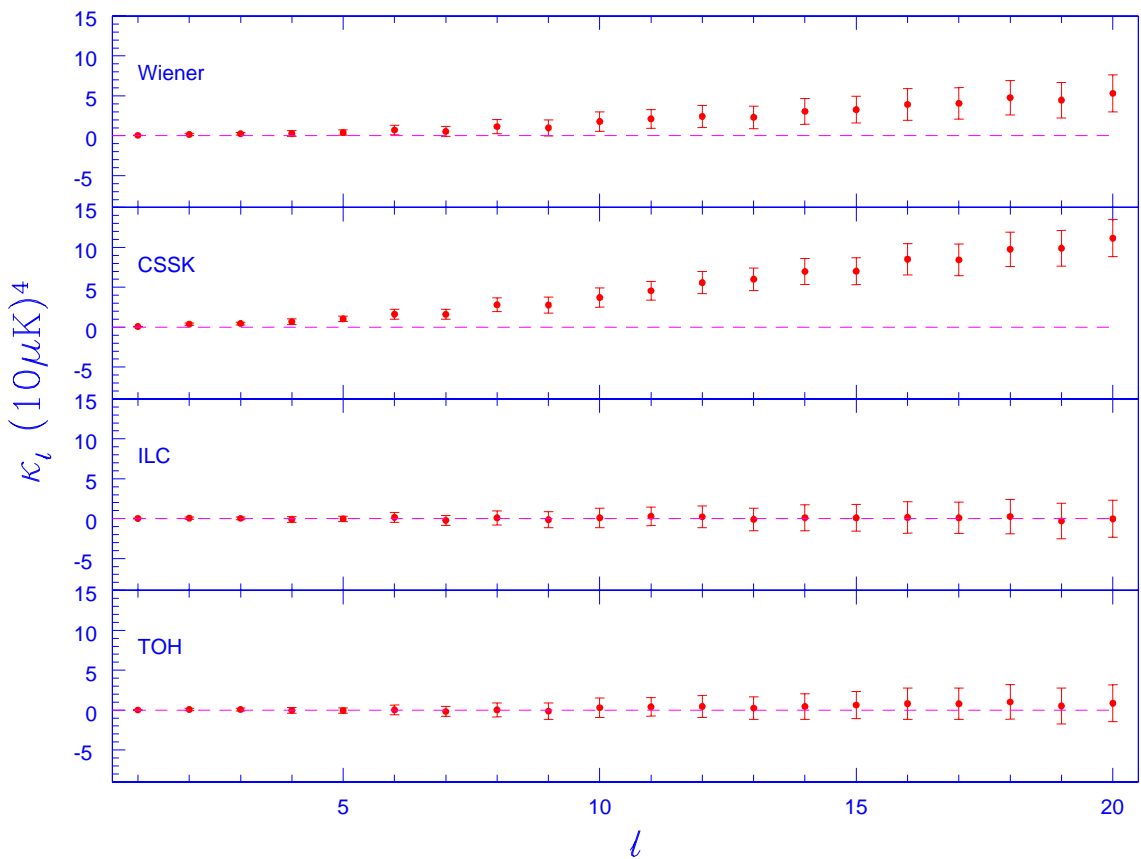}
  \caption{BiPS measured from 4 different WMAP maps filtered by $W_l^S(20,45)$. From bottom to up: TOH  map, ILC map, CSSK foreground free map with a galactic cut, Wiener filtered map. We see that Wiener filtering has the similar effect as galactic cut. (But not exactly the same) }
   \label{wiener}
 \end{figure}

We use the estimator given in eqn.(\ref{estimator}) to measure BiPS for the given CMB maps. We compute the BiPS for all window functions shown in
Fig~\ref{cl_wmap_cl100}. Results for three of these windows are plotted in Figs. \ref{kappa_wmap_400}, \ref{kappa_wmap_3020}, and \ref{wiener}. In the low-$l$ regime, where we have kept the low multipoles, BiPS for all three given maps are consistent with zero (Fig. \ref{kappa_wmap_400}). But in the intermediate-$l$ regime (Fig. \ref{kappa_wmap_3020}), although BiPS of ILC and TOH maps are well consistent with zero, the CSSK map shows a rising tail in BiPS due to the galactic mask. To confirm it, we compute the BiPS for the ILC map with a 10-degree cut around the galactic plane (filtered with the same window function). The result is shown on the top panel of Fig. \ref{kappa_wmap_3020}. Another interesting effect is seen when we apply a $W_l^S(20,45)$ filter, where Wiener filtered map has a non zero BiPS very similar to that of CSSK but weaker (Fig. \ref{wiener}). The reason is that Wiener filter  takes out more modes from regions with more foregrounds since these are inconsistent with the theoretical model. As a result, a Wiener filtered map at $W_l^s(20,45)$ filter has a BiPS similar to a cut sky map. The fact that Wiener map has less power at the Galactic plane can even be seen by eye!  Comparing Fig. \ref{kappa_wmap_3020} to Fig.\ref{wiener} we see that using different filters allows us to uncover different types of violation of SI in a CMB map. In our analysis we have used a set of filters which enables us to probe SI breakdown on angular scales $l\lsim 60$.

\begin{figure}[p]
  \includegraphics[scale=0.8, angle=0]{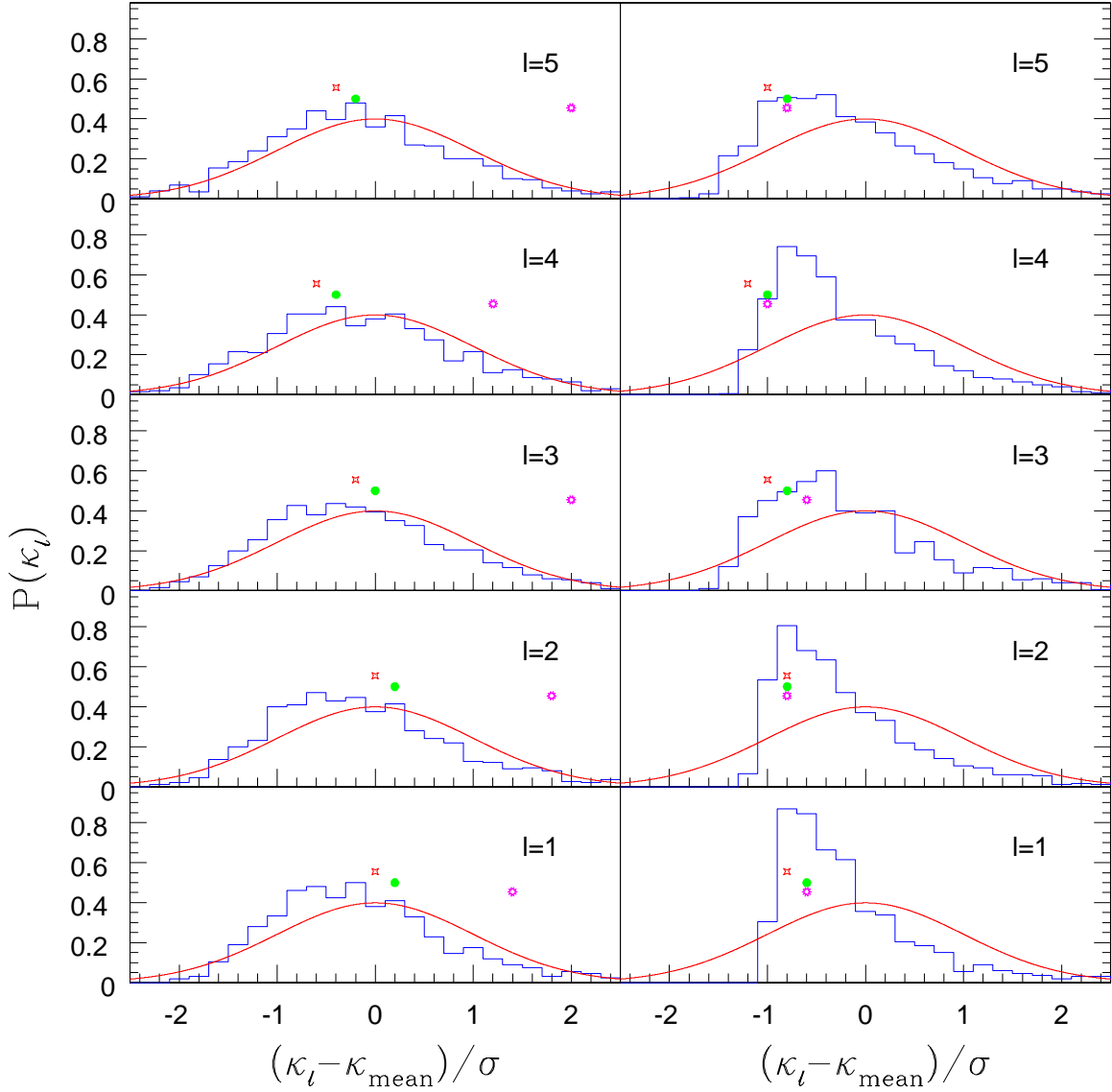}
  \caption{  Probability distribution function for $\kappa_1$
to $\kappa_5$ constructed from $1000$ realizations. The left panel
shows the PDF for the maps filtered with $W^S_l(20,30)$ (left panel)
and $W^G_l(l_s=40)$ (right panel). The latter is more skewed, which
explains the $\sim 1\sigma$ shift in $\kappa_{\ell}$ values shown in
Fig. (\ref{kappa_wmap_400}). The green, magenta and red (circular, pentagonal and rectangular) points represent ILC, CSSK and TOH maps, respectively. The smooth solid curves are Gaussian approximations.}
\label{pdf1to5}
\end{figure}

\begin{figure}[h]
  \includegraphics[scale=0.6, angle=0]{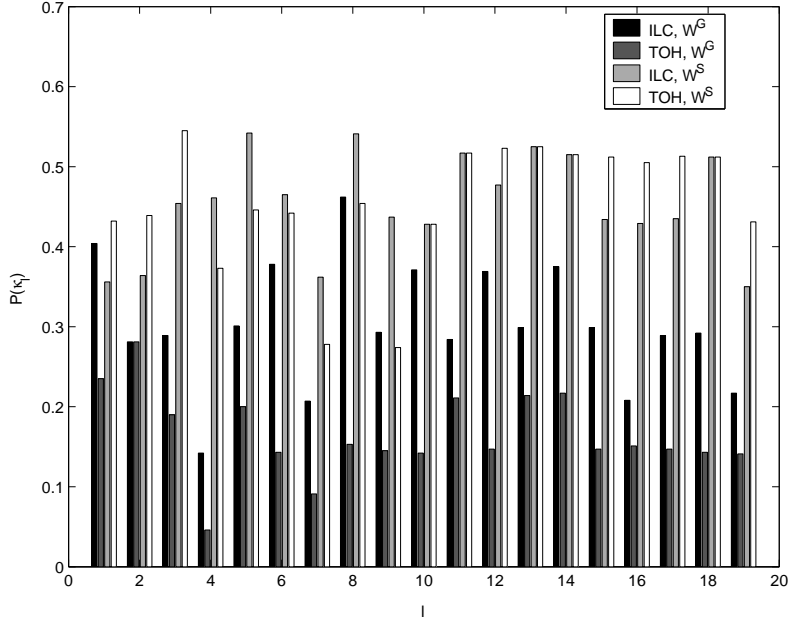}
  \caption{The probability of the two {\it WMAP} based CMB
  maps being  SI when filtered by $W_l^S(20,30)$ 
  and a Gaussian filter $W_l^G(40)$. } 
\label{prob3020}
\label{prob400}
\end{figure}

The BiPS measured from $1000$ simulated SI realizations of 
$C_l^T$ is used to estimate the probability distribution functions
(PDF), $p(\tilde \kappa_{\ell})$. A sample of the PDF for two windows is shown in Fig.~\ref{pdf1to5}. Measured values of BiPS for ILC, TOH and CSSK maps are plotted  on the same plot. BiPS for ILC and TOH maps are located very close to the peak of the PDF. 
We compute the individual probabilities of the map
being SI for each of the measured $\tilde{\kappa}_{\ell}$. This probability is
obtained by integrating the PDF beyond the measured $\tilde
\kappa_{\ell}$. To be precise, we compute 
\bea
\label{probability}
P(\tilde\kappa_{\ell}| {C_l^T})&=& P(\kappa_{\ell} >
\tilde\kappa_{\ell})=\int_{\tilde\kappa_{\ell}}^{\infty} d\kappa_{\ell}
\,p(\kappa_{\ell}), \,\,\,\,\,\tilde\kappa_{\ell} > 0, \\ \nonumber &=&
P(\kappa_{\ell} < \tilde\kappa_{\ell})=\int_{-\infty}^{\tilde\kappa_{\ell}}d\kappa_{\ell}
\,p(\kappa_{\ell}), \,\,\,\,\,\tilde\kappa_{\ell} < 0.  
\eea 
The probabilities obtained are shown in Figs. \ref{prob3020} and 
  \ref{prob400} for $W^S(20,30), W_l^G(40)$ and $W_l^G(4)$. The probabilities for 
the $W^S_l(20,30)$
window function are greater than $0.25$ and the minimum probability at
$\sim 0.05$ occurs at $\kappa_4$ for $W^G(40)$. The reason for systematically lower SI probabilities for $W{_l}^S(20,30)$ as compared to $W{_l}^G(40)$ is simply due to lower cosmic variance of the former. The contribution to the cosmic variance of BiPS is dominated by the low spherical harmonic multipoles. Filters that suppress the $a_{lm}$ at low multipoles have a
lower cosmic variance.

\begin{figure}[h]
  \includegraphics[scale=0.3, angle=-90]{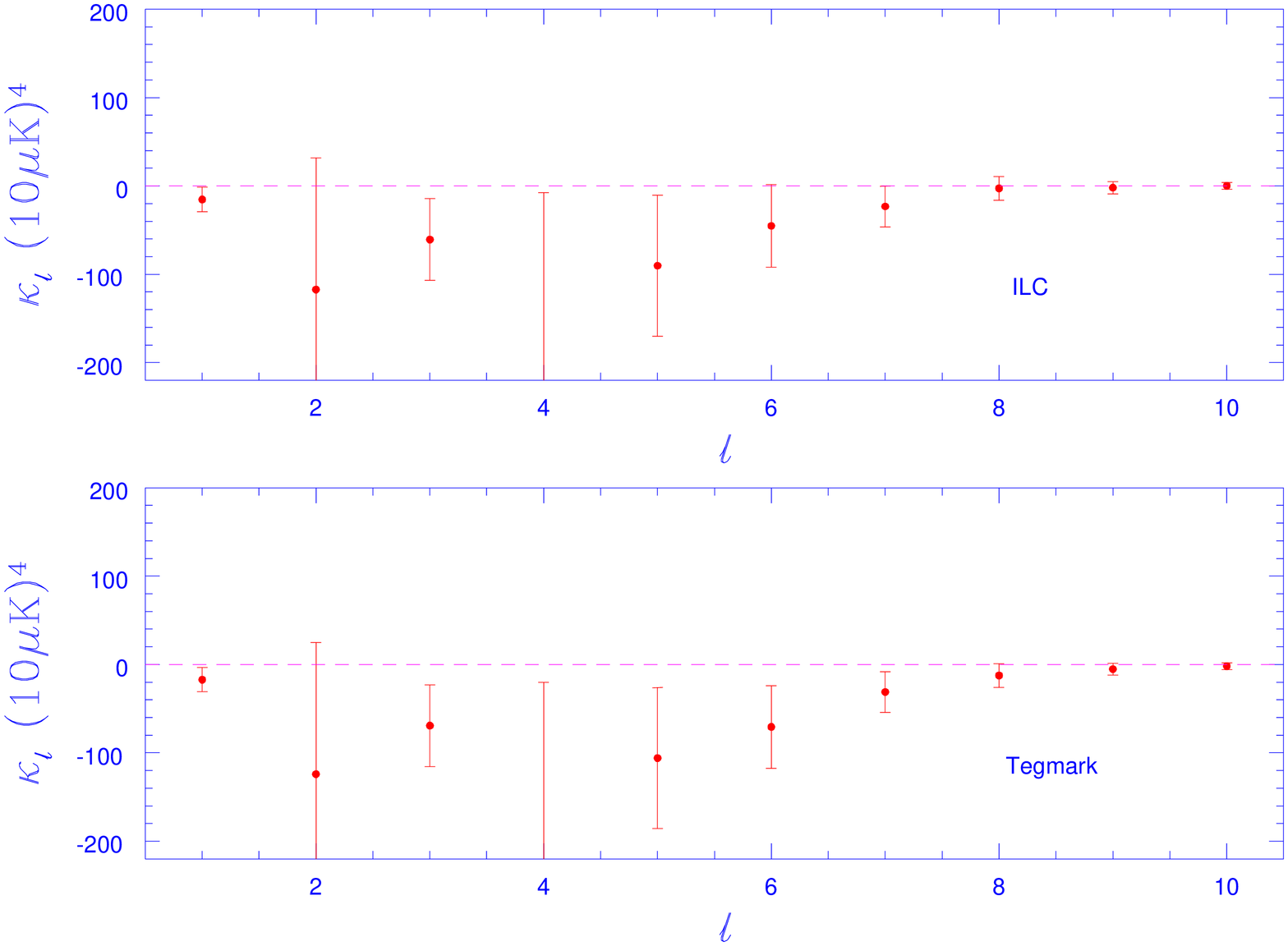}
  \includegraphics[scale=0.3, angle=-90]{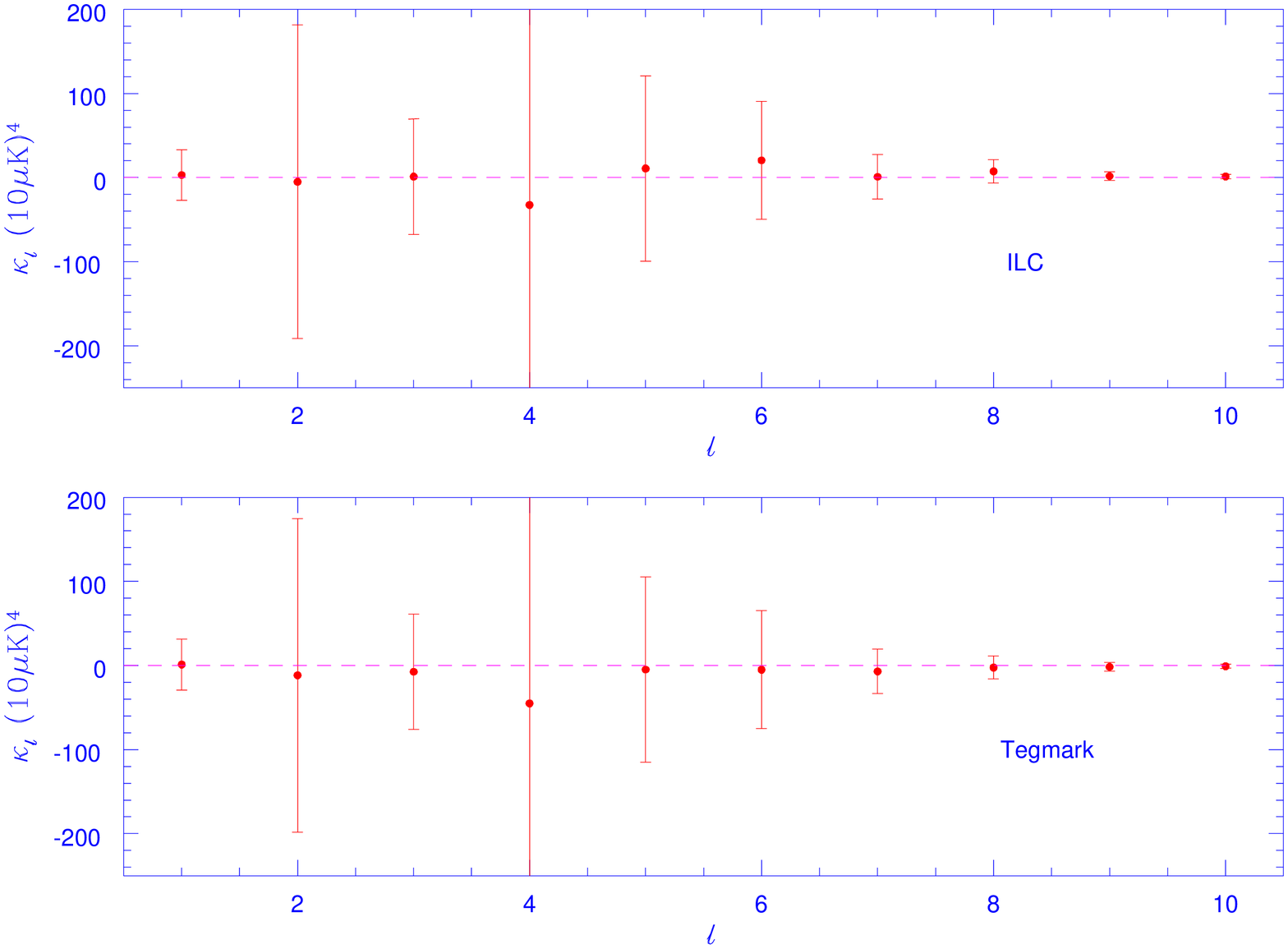}
  \caption{ Figure compares the measured values of $\kappa_\ell$ for
maps A and B filtered to retain power only on the lowest multipoles,
$l=2$ and $l=3$ assuming the WMAP theoretical spectrum WMAPbf ({\it left})
and a model spectrum that matches the suppressed power at the lowest
multipoles \cite{shaf_sour04}. The non zero $\kappa_{\ell}$ `detections'
assuming the WMAP theoretical spectrum become consistent with zero for
a $C_l^T$ that has power suppressed at low multipoles.}
\label{kappa_wmap_0_4}
\end{figure}

It is important to note that the above probability is a conditional
probability of measured $\tilde\kappa_{\ell}$ being SI given the
theoretical spectrum $C_l^T$ (used to estimate the bias).  A final
probability emerges as the Bayesian chain product with the probability
of the theoretical $C_l^T$ used given data. Hence, small difference in
these conditional probabilities for the two maps are perhaps not
necessarily significant. Since the BiPS is close to zero, the
computation of a probability marginalized over the $C_l^T$ may be
possible using Gaussian (or, improved) approximation to the PDF of
$\kappa_{\ell}$. The important role played by the choice of the theoretical
model for the BiPS measurement is shown for a $W_l$ that retains power
in the lowest multipoles, $l=2$ and $l=3$. Assuming  $C_l^T$,
there are hints of non-SI detections in the low $\ell$'s (left panel
of Fig.~\ref{kappa_wmap_0_4}). We also compute the BiPS using a
$C_l^T$ for a model that accounts for suppressed quadrupole and
octopole in the WMAP data~\cite{shaf_sour04}. The mild detections of a
non zero BiPS vanish for this case (right panel of Fig.~\ref{kappa_wmap_0_4}).

\section{Discussion and Conclusion}

The SI of the CMB anisotropy has been under scrutiny after the release of the first year of WMAP data. We use the BiPS which is sensitive to structures and patterns in the underlying total two-point correlation function as a statistical tool of searching for departures from SI.   We carry out a BiPS analysis of WMAP full
sky maps. We find no strong evidence for SI violation in the WMAP CMB
anisotropy maps considered here. We have verified that our null
results are consistent with measurements on simulated SI maps.  The
BiPS measurement reported here is a Bayesian estimate of the
conditional probability of SI (for each $\kappa_{\ell}$ of the BiPS) given
an underlying theoretical spectrum $C_l^T$. We point out that the
excess power in the  $C_l^T$ with respect to the measured $C_l$
from WMAP at the lowest multipoles tends to indicate mild deviations
from SI. BiPS measurements are shown to be consistent with SI assuming
an alternate model $C_l^T$ that is consistent with suppressed power on
low multipoles. Note that it is possible to band together $\kappa_{\ell}$
measurements to tighten the  error bars further. The full sky maps and the restriction to
low $l<60$ (where instrumental noise is sub-dominant) permits the use
of our analytical bias subtraction and error estimates. The excellent
match with the results from numerical simulations is a strong
verification of the numerical technique.  This is an important check
before using Monte-Carlo simulations in future work for computing BiPS
from CMB anisotropy sky maps with a galactic mask and non uniform
noise matrix.

There are strong theoretical motivations for hunting for SI
violation in the CMB anisotropy. The possibility of non-trivial cosmic
topology is a theoretically well motivated possibility that has also
been observationally targeted~\cite{ell71, lac_lum95, lev02, linde}.  The
breakdown of statistical homogeneity and isotropy of cosmic
perturbations is a generic feature of ultra large scale structure of
the cosmos, in particular, of non trivial cosmic topology \cite{bps}.
The underlying correlation patterns in the CMB anisotropy in a
multiply connected universe is related to the symmetry of the
Dirichlet domain. The BiPS expected in flat,
toroidal models of the universe has been computed and shown to be
related to the principle directions in the Dirichlet
domain \cite{us_prl}. As a tool for constraining cosmic topology, the
BiPS has the advantage of being independent of the
overall orientation of the Dirichlet domain with respect to the
sky. Hence, the null result of BiPS can have important implication for
cosmic topology. This approach complements direct search for signature
of cosmic topology~\cite{circles, staro} and our results are consistent
with the absence of the matched circles and the null S-map test of the
WMAP CMB maps~\cite{circles04, angelwmap}. Full Bayesian likelihood
comparison to the data of specific cosmic topology models is another
approach that has applied to COBE-DMR data~\cite{bps}. Work is in
progress to carry out similar analysis on the large angle WMAP data.
We defer to future publication, detailed analyzes and constraints on
cosmic topology using null BiPS measurements, and the comparison to
the results from complementary approaches. There are also other
theoretical scenarios that predict breakdown of SI that can be 
probed using BiPS, e.g., primordial
cosmological magnetic fields \cite{DKY, gang}.

The null BiPS results also has implications for the observation and data analysis techniques used to create the CMB anisotropy maps. Observational artifacts such as  non-circular beam, inhomogeneous noise correlation, residual stripping patterns, etc.  are potential sources of SI breakdown.  Our null BiPS results confirm that these artifacts do not significantly contribute to the maps studied here. Foreground residuals can also be sources of SI breakdown. The extent to which BiPS probes foreground residuals is yet to be fully studied and explored. We do not see any significant effect of the residual foregrounds in ILC and the TOH maps as it was mentioned by \cite{erik04c}. This can not be necessarily called a discrepancy between the two results unless we know what should have been seen in the BiPS. The question is if the signal is strong enough and whether the effect smeared out in bipolar multipole space within our angular $l$-space window. On the other hand, the very fact that BiPS does show a strong signal for the Wiener filtered map, mean that at some level BiPS is sensitive to galactic residuals. 

In summary, we study the Bipolar
power spectrum (BiPS) of CMB which is a promising measure of SI. 
We find null measurements of the BiPS for a selection of
full sky  CMB anisotropy maps based on the first year of WMAP
data. Our results rule out radical violation of statistical isotropy,
and are consistent with null results for matched circles and the S-map
tests of SI violation.  Pending a more careful comparison, the results
do not necessarily conflict with a number of other statistical tests, that hint at SI violation at low to modest statistical significance. 

\acknowledgments 
The authors are very thankful for close interaction with David
Spergel, Glenn Starkman and Neil Cornish. We also
acknowledge useful discussions with Dick Bond, Dmitry Pogosyan and
Carlo Contaldi.  AH thanks Istvan Szapudi, Simon Prunet, Uros Seljak and Jacques Delabrouille for illuminating discussions.  Use of the HPC facility of IUCAA is acknowledged.

\appendix
\section{More on Gaussianity}
\label{AppGauss}
The temperature anisotropy at a given space-time point $({\mathbf x},\tau)$ can be written as a superposition of plane waves in $k$-space
\be
\Delta T({\mathbf x},\hat{n},\tau)=\int{\mathrm{d}^3k \, e^{i{\mathbf k} \cdot {\mathbf x}} \, \Delta T({\mathbf k},\hat{n},\tau)}.
\ee
As discussed in \cite{Ma95}, the dependence on $\hat{n}$ arises only through ${\mathbf k}\cdot\hat{n}$. Therefore $\Delta T({\mathbf k},\hat{n},\tau)$ may be represented as
\be
 \Delta T({\mathbf k},\hat{n},\tau)=\sum_{l=0}^{\infty}{(-i)^l(2l+1)\Delta T_l({\mathbf k}, \tau)P_l({\mathbf k}\cdot\hat{n})}.
\ee
The anisotropy coefficients $\Delta T_l({\mathbf k}, \tau)$ are random variables whose amplitudes and phases depend on the initial perturbations and can be written as
\be 
 \Delta T_l({\mathbf k}, \tau)=\phi_0({\mathbf k}) \Delta T_l(k, \tau),
\ee
where $\phi_0({\mathbf k})$ is the initial potential perturbation and $ \Delta T_l(k, \tau)$ is the photon transfer function, {\it i.e.} solution of Boltzmann equation with $\phi_0({\mathbf k})=1$. In general,  $\Delta T_l(k, \tau)$ is given by  the integral solution 
\be
\Delta T_l(\beta)=\int_{0}^{\tau_0}{\mathrm{d} \tau \Phi_{\beta}^{l}(\tau_0-\tau)S(\beta, \tau)}, 
\ee
in which $S(\beta, \tau)$ is the source function, $\Phi_{\beta}^{l}$ is the ultra-spherical Bessel function and $\beta=k-K$, where $K=H_0^2(\Omega_0-1)$ is the curvature \cite{zaldar98}. In the flat case, the above expression has a very simple form
\be
\Delta T(k)=\int_{0}^{\tau_0}{\mathrm{d} \tau \,e^{i\frac{{\mathbf k}\cdot\hat{n}}{k}(\tau-\tau_0)} S(k,\tau)},
\ee
and hence, the temperature anisotropies are given by
\be
\label{deltat}
\Delta T({\mathbf x},\hat{n})=\int{\mathrm{d}^3k \, e^{i{\mathbf k} \cdot {\mathbf x}} \,  \phi_0({\mathbf k}) [\int_{0}^{\tau_0}{\mathrm{d} \tau \,e^{i\frac{{\mathbf k}\cdot\hat{n}}{k}(\tau-\tau_0)} S(k,\tau)}]}.
\ee
Equation (\ref{deltat}) clearly shows that Gaussian perturbations in primordial gravitational potential result a Gaussian CMB temperature anisotropy field.

\section{SI condition in harmonic space}
\label{moreharmonic}
We substitute the spherical harmonic coefficients, $a_{lm}$, from eqn. (\ref{alm}) in the covariance matrix to express it in terms of the two point correlation
\bea
\langle a_{lm}a^*_{l^\prime m^\prime}\rangle
 & = &  \int{\int{\mathrm{d} \Omega_{\hat{n}} \mathrm{d}\Omega_{\hat{n}'}Y_{lm}^{*}(\hat{n})Y_{l'm'}(\hat{n}')\langle\Delta T(\hat{n})} \Delta T(\hat{n}')}\rangle
\\ \nonumber 
 & = &  \int{\int{\mathrm{d} \Omega_{\hat{n}} \mathrm{d}\Omega_{\hat{n}'}Y_{lm}^{*}(\hat{n})Y_{l'm'}(\hat{n}') C(\hat{n}, \hat{n}')}}.
\eea
Under the condition of SI, eqn. (\ref{statiso}), and after making use of the spherical harminc expansion of Legendre polynomials, eqn. (\ref{Legexpansion}), in  eqn. (\ref{LegendreTransform}), the above expression will become
\bea
\langle a_{lm}a^*_{l^\prime m^\prime}\rangle
 & = & \int{\int{\mathrm{d} \Omega_{\hat{n}} \mathrm{d}\Omega_{\hat{n}'}Y_{lm}^{*}(\hat{n})Y_{lm}(\hat{n}') \sum_{l_1}{C_{l_1}\sum_{m_1}{Y_{l_1m_1}(\hat{n})Y_{l_1m_1}(\hat{n}')}}}}
\\ \nonumber 
 & = & \sum_{l_1}{C_{l_1}\sum_{m_1}{\delta_{ll_1}\delta_{l'l_1}\delta_{mm_1}\delta_{m'm_1}}} 
\\ \nonumber 
 & = & C_{l} \delta_{ll^\prime}\delta_{mm^\prime}, 
\eea
which is the SI condition in harmonic space, eqn. (\ref{diagonalcovmat}), and in the last line, the orthonormality relation of spherical harmonics, eqn. (\ref{A2}) has been used. 

\section{ $A^{\ell M}_{l_1 l_2}$ are linear combinations of  $\langle a_{lm} a^*_{l^\prime m^\prime}\rangle$}
 \label{ALMproperties}
 The BiPoSH coefficients, $A^{\ell M}_{l_1 l_2}$, are linear combinations $a_{lm}$. To see that, in the definition of two point correlation 
 we can substitute for $\Delta T/T$ from eqn. (\ref{yelemexpand}), which will lead to the expansion of $C(\hat{n}_1, \hat{n}_2)$ in terms of spherical harmonics
 \bea
 C(\hat{n}_1, \hat{n}_2)\,&=&\,\langle \Delta T(\hat{n}_1)\Delta T(\hat{n}_2)\rangle
 \\ \nonumber
 \,&=&\,\sum_{lm}\sum_{l'm'}\langle a_{lm}a^{*}_{l'm'}\rangle Y_{lm}(\hat{n}_1)
 Y^{*}_{l'm'}(\hat{n}_2). 
 \eea
 This can be substituted in the definition of $A^{\ell M}_{l_1 l_2}$ and eqn. (\ref{alml1l2}) can be written as
 \bea
 A^{\ell M}_{l_1 l_2}\,&=&\, \int d\Omega_1 \int d\Omega_2 C(\hat{n}_1, \hat{n}_2) \{ {\boldmath{Y}}_{l_1}(\hat{n}_1)\times Y_{l_2}(\hat{n}_2)\}_{\ell M}^{*}
 \\ \nonumber
 \,&=&\, \int d\Omega_1 \int d\Omega_2 \sum_{lm}\sum_{l'm'}\langle a_{lm}a^{*}_{l'm'}\rangle Y_{lm}(\hat{n}_1) Y^{*}_{l'm'}(\hat{n}_2) \sum_{m_1m_2}C^{LM}_{l_1m_1l_2m_2} Y^{*}_{l_1m_1}(\hat{n}_1)Y^{*}_{l_2m_2}(\hat{n}_2)
 \\ \nonumber
 \,&=&\, \int d\Omega_1 \int d\Omega_2 \sum_{lm}\sum_{l'm'}\langle a_{lm}a^{*}_{l'm'}\rangle Y_{lm}(\hat{n}_1) Y^{*}_{l'm'}(\hat{n}_2) \sum_{m_1m_2}C^{LM}_{l_1m_1l_2m_2} Y^{*}_{l_1m_1}(\hat{n}_1) (-1)^{m_2} Y_{l_2 -m_2}(\hat{n}_2)
 \\ \nonumber
 \,&=&\,\sum_{m_1m_2}(-1)^{m_2}\langle a_{l_1m_1}a^{*}_{l_2 -m_2}\rangle C^{LM}_{l_1m_1l_2m_2}
 \eea
 By changing the dummy variable $m_2 \to -m_2$ we will get
 \be \label{ALMvsalmapp}
 A^{\ell M}_{l_1 l_2}\,=\, \sum_{m_1m_2} \langle a_{l_1m_1}a^{*}_{l_2 m_2}\rangle (-1)^{m_2} C^{LM}_{l_1m_1l_2 -m_2}.
 \ee
 
 \section{Simplifying the real-space expression for BiPS}
\label{simplification}
The BiPS in real-space is given by the following expression.
\be \label{kappal2}
\kappa_{\ell}\, = \, (\frac{2\ell+1}{8\pi^2})^2 \int d\Omega_1 
\int d\Omega_2 \int dR_1 \chi_{\ell}(R_1) \int dR_2 \chi_{\ell}(R_2) 
C(R_1\hat{q}_1,R_1\hat{q}_2)C(R_2\hat{q}_1,R_2\hat{q}_2)
\ee

Now if we change the variables in the following way
\bea
R_1\hat{q}_1=\hat{q}_1 \\ \nonumber
R_1\hat{q}_2=\hat{q}_2
\eea
We will have
\be
\kappa_{\ell}\, = \, (\frac{2\ell+1}{8\pi^2})^2 \int d\Omega_1 
\int d\Omega_2 \int dR_1 \chi_{\ell}(R_1) \int dR_2 \chi_{\ell}(R_2) 
C(\hat{q}_1,\hat{q}_2)C(R_2R_1^{-1}\hat{q}_1,R_2R_1^{-1}\hat{q}_2),
\ee
and with 
\be
R=R_2R_1^{-1}
\ee
we will get
\be
\kappa_{\ell}\, = \, (\frac{2\ell+1}{8\pi^2})^2 \int d\Omega_1 
\int d\Omega_2 \int dR_1 \chi_{\ell}(R_1) \int dR \chi_{\ell}(RR_1) 
C(\hat{q}_1,\hat{q}_2)C(R\hat{q}_1,R\hat{q}_2).
\ee
Using eqn. (\ref{4cosvar}) we can write
\bea
\int dR_1 \chi_{\ell}(R_1) \chi_{\ell}(RR_1) \,&=&\, \sum_{m_1}\sum_{mm'}
\int dR_1 D_{m_1m_1}^{*\ell}(R_1)D_{mm'}^{\ell}(R)D_{m'm}^{\ell}(R_1)
\\ \nonumber
\,&=&\, \sum_{m_1}\sum_{mm'} D_{mm'}^{\ell}(R) \frac{8\pi^2}{2\ell+1} 
\delta_{m_1m'} \delta_{m_1m}
\\ \nonumber
\,&=&\, \frac{8\pi^2}{2\ell+1} \sum_{m_1}  D_{m_1m_1}^{\ell}(R)
\\ \nonumber
\,&=&\, \frac{8\pi^2}{2\ell+1} \chi_{\ell}(R).
\eea
So, the $\kappa_{\ell}$ can be written in this form:
\be
\label{simplified}
\kappa_{\ell}\, = \, \frac{2\ell+1}{8\pi^2} \int d\Omega_1 
\int d\Omega_2 C(\hat{q}_1,\hat{q}_2)  \int dR \chi_{\ell}(R) 
C(R\hat{q}_1,R\hat{q}_2).
\ee
\section{More on Cosmic Variance}
\label{AppCosVar}
The cosmic variance, $\langle \tilde{\kappa}_{\ell}^2\rangle-\langle\tilde{\kappa}_{\ell}\rangle^2$, has 96 terms similar to the one shown in eqn. (\ref{firstterm}) but with different combinations of $R$'s and $\hat{n}$'s. Among them, 40 terms contain at least one term like $C(R\hat{n}_1,R\hat{n}_2)$. These terms only contribute to $\kappa_0$ because when SI holds,  $C(R\hat{n}_1,R\hat{n}_2)= C(\hat{n}_1,\hat{n}_2)$ and hence the integral in eqn. (\ref{simplified}) will be proportional to $\delta_{\ell 0}$, which is not in the interest of us. So, we can ignore them and we will be left with 56 terms. These terms, depending on the number of connected correlations in them, can be classified into 4 major groups:
\begin{enumerate}
\item One-loop terms, such as
\be
  C(\hat{n}_1,R'\hat{n}_3) C(\hat{n}_3,R\hat{n}_2)C(\hat{n}_2,R'\hat{n}_4)C(\hat{n}_4,R\hat{n}_1).
\ee
These terms get reduced to expressions with a $C_l^4$ factor and 4 Clebsch-Gordan coefficients (summation symbols are omitted for brevity),
\be
C_{l_1}^4C^{\ell M}_{l_{1}-m_{1}l_{1}-m_{3}}C^{\ell M}_{l_{1}m_{ 5}l_{1}m_{ 7}}C^{\ell M'}_{l_{1}m_{ 7}l_{1}m_{ 5}}C^{\ell M'}_{l_{1}-m_{1}l_{1}-m_{3}}.
\ee
There are 24 terms like this but with different combinations of indices.
\item Two-loop terms -- type I, such as
\be
C(\hat{n}_1,\hat{n}_3)C(R'\hat{n}_3,R\hat{n}_2)C(\hat{n}_2,R\hat{n}_1)C(\hat{n}_4,R'\hat{n_4}).
\ee
These terms get reduced to expressions with a $C_{l_1}^3C_{l_2}$ factor and 4 Clebsch-Gordan coefficients
\be
C_{l_1}^3 C_{l_7}C^{\ell M}_{l_{1}-m_{1}l_{1}-m_{3}}C^{\ell M}_{l_{1}-m_{3}l_{1}m_{ 5}}C^{\ell M'}_{l_{1}m_{ 1}l_{7}-m_{7}}C^{\ell M'}_{l_{1}-m_{5}l_{7}-m_{7}}.
\ee
There are 16 terms like this but with different combinations of indices.
\item Two-loop terms -- type II, such as
\be
 C(\hat{n}_1,R'\hat{n}_4)C(\hat{n}_4,R\hat{n}_1)C(\hat{n}_2,R'\hat{n}_3)C(R\hat{n}_2,\hat{n}_3).
\ee
These terms get reduced to expressions with a $C_{l_1}^2C_{l_2}^2$ factor and 4 Clebsch-Gordan coefficients
\be
C_{l_1}^2  C_{l_3}^2 C^{\ell M}_{l_{1}-m_{1}l_{3}-m_{3}}C^{\ell M}_{l_{1}m_{ 5}l_{3}m_{ 7}}C^{\ell M'}_{l_{1}m_{ 5}l_{3}m_{ 7}}C^{\ell M'}_{l_{1}-m_{1}l_{3}-m_{3}}.
\ee
There are 8 terms like this but with different combinations of indices.
\item Three-loop terms, such as
\be
C(\hat{n}_1,\hat{n}_3) C(R'\hat{n}_3,R\hat{n}_1)C(\hat{n}_2,R\hat{n}_2)C(\hat{n}_4,R'\hat{n}_4).
\ee
These terms get reduced to expressions with a $C_{l_1}^2C_{l_2}C_{l_3}$ factor and 4 Clebsch-Gordan coefficients
\be
C_{l_1}^2  C_{l_3} C_{l_7}C^{\ell M}_{l_{3}-m_{3}l_{1}-m_{1}}C^{\ell M}_{l_{3}-m_{3}l_{1}m_{ 5}}C^{\ell M'}_{l_{1}m_{ 1}l_{7}-m_{7}}C^{\ell M'}_{l_{1}-m_{5}l_{7}-m_{7}}.
\ee
There are 8 terms like this but with different combinations of indices.
\end{enumerate}

Using symmetry properties of Clebsch-Gordan coefficients, eqn. (\ref{CGsymmetries}), and their summation rules, eqn. (\ref{A10}), it is possible to simplify these 56 terms immensely. After tedious algebra we would obtain the final expression for the kernel of cosmic variance of the bipolar power spectrum:
\bea
& 4 &\,C_{l_1}^4 [2\frac{(2\ell+1)^2}{2l_1+1}+(-1)^{\ell}(2\ell+1)+(1+2(-1)^{\ell})F_{ll}^{\ell}]\,\{l_1l_1\ell\}+   \nonumber \\ 
& 4 &\, C_{l_1}^2  C_{l_2}^2 [(2\ell+1)+F_{l_1l_2}^{\ell}]\,\{l_1l_2\ell\} +   \nonumber \\ 
& 8 &\, C_{l_1}^2  C_{l_2} C_{l_3}\frac{(2\ell+1)^2}{2l_1+1}\,\{l_1l_2\ell\} \,\{l_1l_3\ell\}  +   \nonumber \\ 
& 16 &\,(-1)^{\ell}\,C_{l_1}^3 C_{l_3}\frac{(2\ell+1)^2}{2l_1+1}\,\{l_1l_1\ell\} \,\{l_1l_3\ell\}  
\eea
where the $3j$-symbol $\{l_1l_2\ell\}$ is given by
\begin{displaymath}
\{l_1l_2\ell\} = \left\{
\begin{array}{ll}
1 &\textrm{if $l_1+l_2+\ell$ is integer and $|l_1-l_2|\leq \ell \leq (l_1+l_2)$,}  \\
0 &\textrm{otherwise,} 
\end{array} \right.
\end{displaymath}
and $F_{l_1l_3}^\ell$ is defined as
\be
 F_{l_1l_3}^\ell = \!\!\!\!
\sum_{m_1m_2=-l_1}^{l_1}\,\sum_{m_3m_4=-l_3}^{l_3}
\sum_{M,M'=-\ell}^\ell C^{\ell M}_{l_{1}-m_{1}l_{3}-m_{3}}C^{\ell
M}_{l_{1}m_{2}l_{3}m_{4}} C^{\ell M'}_{l_{3}m_{4}l_{1}m_{ 1}}C^{\ell
M'}_{l_{3}-m_{3}l_{1}-m_{2}}.
\ee
This should be summed over all indices except $\ell$ to result the analytical expression for the cosmic variance given in eqn. (\ref{klcv}).
\section{Useful Standard Integrals and Summation Rules}\label{appA}
For completeness we list a set of standard integrals and summation rules which are needed in BiPS calculations. Orthonormality of spherical harmonics
\begin{eqnarray}
 \int d\Omega_{\hat{n}}\, Y_{\ell m}(\hat{n})&=&\sqrt{4\pi}\delta_{\ell0}\delta_{m0},\label{A1}\\
 \int d\Omega_{\hat{n}}\, Y_{l_1m_1}(\hat{n})Y_{l_2m_2}^*(\hat{n})&=&\delta_{l_1l_2}\delta_{m_1m_2},\label{A2}
\end{eqnarray}
\begin{eqnarray}
 \int d\Omega_{\hat{n}} Y_{l_1m_1}(\hat{n})Y_{l_2m_2}(\hat{n})Y_{\ell_3m_3}^*(\hat{n})&=&
 \sqrt{\frac{(2l_1+1)(2l_2+1)}{4\pi(2\ell_3+1)}}
 C_{l_10l_20}^{\ell_30}
 C_{l_1m_1l_2m_2}^{\ell_3m_3}.
 \label{A6}\\
\end{eqnarray}
Symmetry property of spherical harmonics
\be
Y_{lm}^*(\hat{n})=(-1)^mY_{l-m}(\hat{n}).
\ee
Spherical harmonic expansion of Legendre polynomials 
\be  \label{Legexpansion}
P_l(\hat{n}\cdot \hat{n}')=\frac{4\pi}{2l+1}\sum_{m=-l}^{l}{Y_{lm}^*(\hat{n})Y_{lm}( \hat{n}')}.
\ee
Orthonormality relation of $\chi^{\ell}(\omega)$ 
 \be
 \int_{0}^{\pi} \chi^{\ell_1}(\omega) \chi^{\ell_2}(\omega) \sin^2{\omega \over{2}} d\omega
 \,=\, \frac{\pi}{2} \delta_{\ell_1 \ell_2}.
\label{A8} 
\ee
Orthonormality of  $D_{MM}^{\ell}({\mathcal R})$ 
\be 
 \int d{\mathcal R} D_{MM'}^{*\ell}({\mathcal R})D_{mm'}^{l}({\mathcal R})\,=\,\frac{8\pi^2}{2l+1} \, \delta_{l\ell}\delta_{mM}\delta_{m'M'},
\label{A9}
 \ee
\be
\label{4cosvar}
\sum_{mm'}D_{mm'}^{\ell}(R_1)D_{m'm}^{\ell}(R_2)=\chi_{\ell}(R_1R_2)
\ee
Symmetry properties of Clebsch-Gordan coefficients
\bea 
\label{CGsymmetries}
C_{a \alpha b \beta}^{c \gamma} &=& (-1)^{a+b-c} C_{ b \beta a \alpha }^{c \gamma} ,
\\ \nonumber
C_{a \alpha b \beta}^{c \gamma} &=& (-1)^{a+b-c} C_{a -\alpha b -\beta}^{c -\gamma}.
\eea
Summation rules of Clebsch-Gordan coefficients
\bea
\sum_{\alpha \beta} C_{a \alpha b \beta}^{c \gamma} C_{a \alpha b \beta}^{c' \gamma'} &=& \delta_{cc'} \delta_{\gamma \gamma'} \{abc\}\{abc'\}
\nonumber \\
\sum_{a \gamma} C_{a \alpha b \beta}^{c \gamma} C_{a \alpha b' \beta'}^{c \gamma} &=& \frac{2c+1}{2b+1} \delta_{bb'} \delta_{\beta \beta'}\{abc\}\{ab'c\}
\nonumber \\
\sum_{c\gamma} C_{a \alpha b \beta}^{c \gamma} C_{a \alpha' b \beta'}^{c \gamma} &=& \delta_{\alpha \alpha'}\delta_{\beta \beta'}\{abc\}
\label{A10}
\eea


\newpage

\begin{figure}[h]
  \includegraphics[scale=0.5, angle=0]{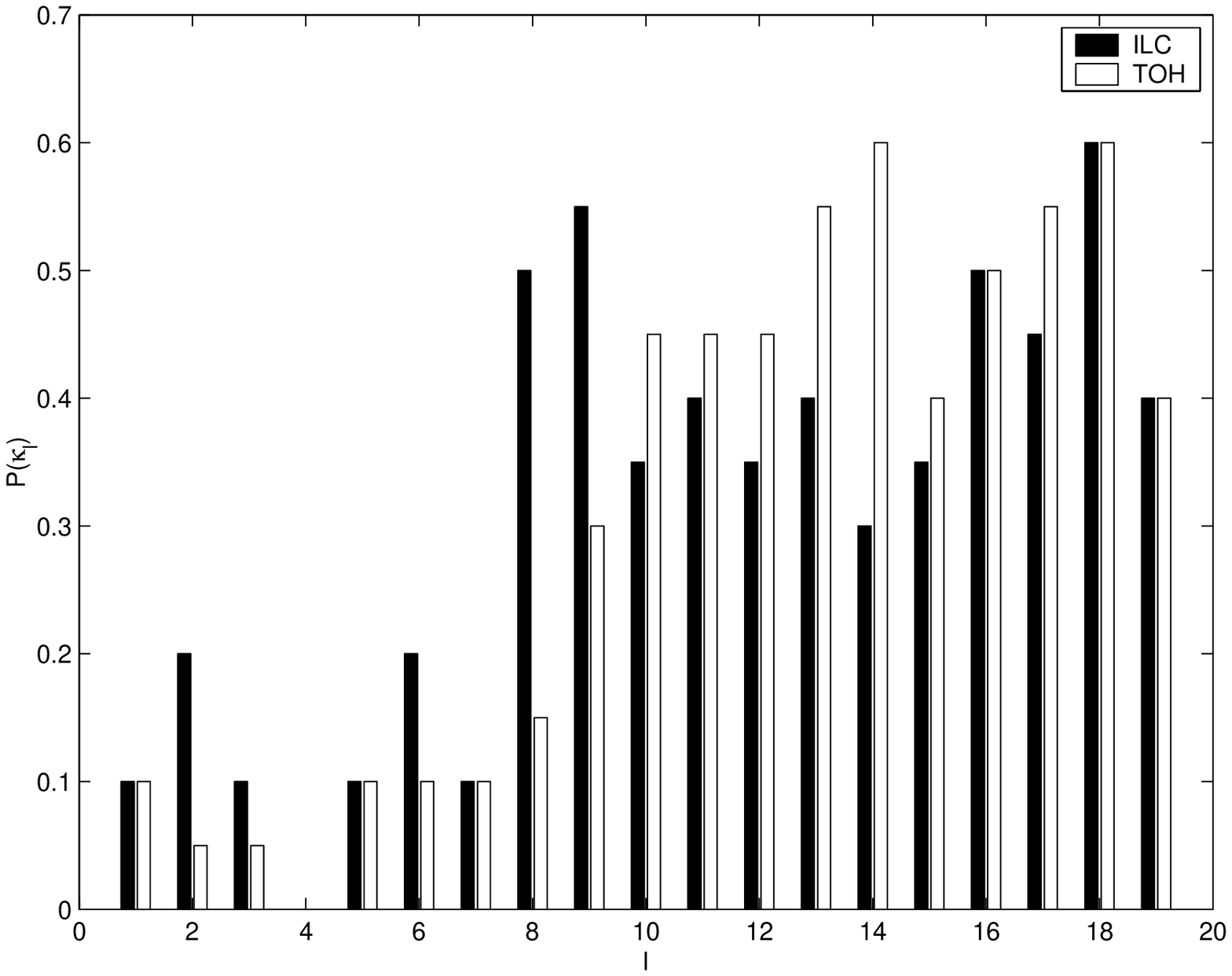}
  \includegraphics[scale=0.5, angle=0]{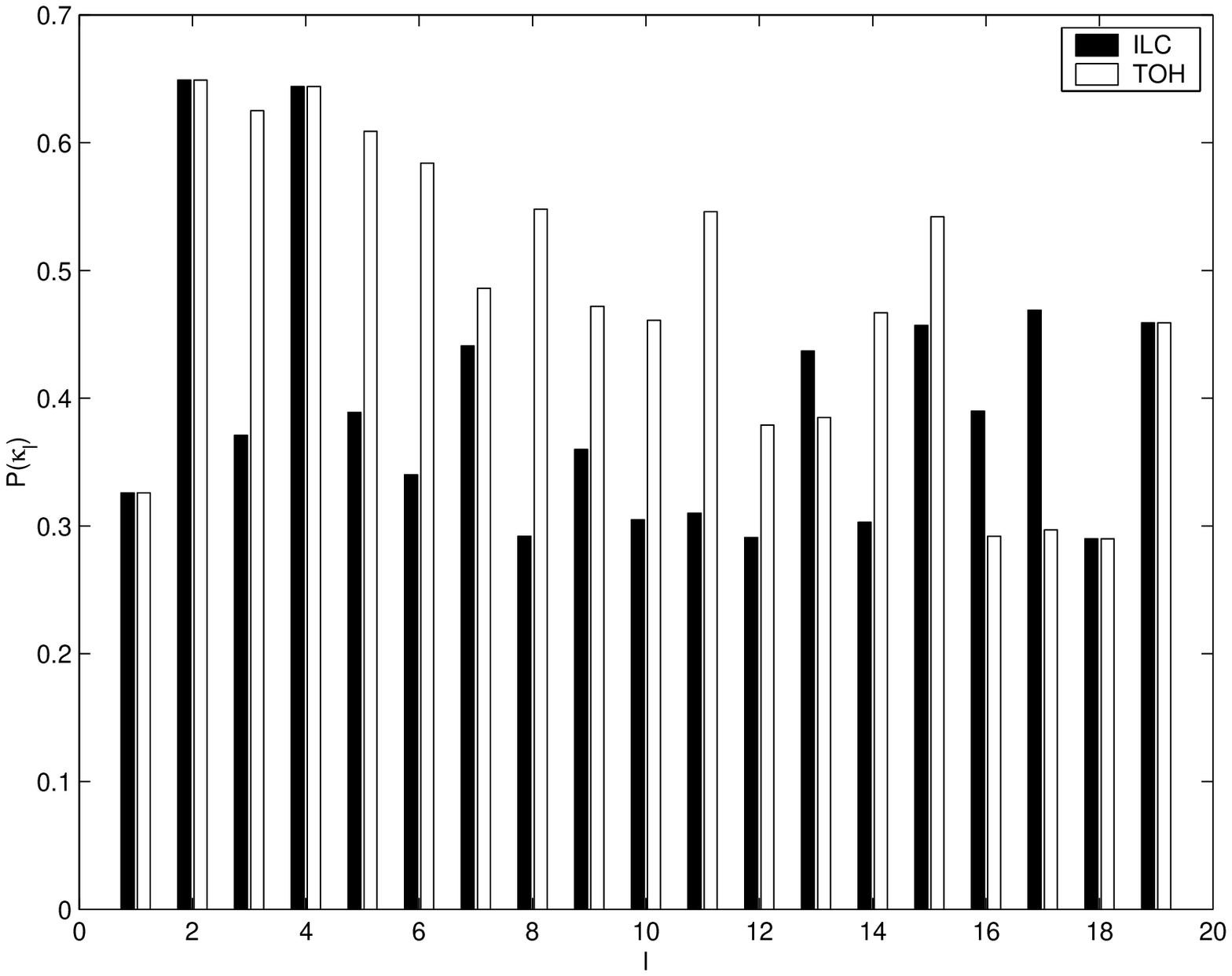}
  \caption{probability of the two {\it WMAP} based CMB maps being SI when filtered by $W_l^G(4)$ to only keep the lowest multipoles when assuming the WMAP theoretical spectrum, $C_l^T$ ({\it{left}}) and  a model spectrum that matches the suppressed power at the lowest multipoles ({\it right}).}
\label{prob400}
\end{figure}


\begin{thebibliography}{99}

\bibitem[Bamba et al. 2004]{bamba2004} Bamba, K.,  Yokoyama, J.,  Phys.Rev. D69 (2004) 043507.

\bibitem[Bardeen et al. 1983]{bar83} Bardeen, J. M., Steinhardt, P. J. \& Turner, M. S. 1983,
Phys. Rev. D 28, 679.

\bibitem[Bartolo et al. 2004]{Bartolo} Bartolo, N., Komatsu, E., Matarrese, S., Riotto, A., {\it preprint} (astro-ph/0406398)

\bibitem[Bennett et al. 2003]{wmap} Bennett,  C. L., et.al., 2003, Astrophys. J. Suppl.,
{\bf 148}, 1.

\bibitem[Bond 2004]{bon04} Bond, J. R. et al. 2004,
Int. J. Theor. Phys. 2004, ed. Verdaguer, E., "The Early Universe:
Confronting theory with observations" (June 21-27, 2003)
(astro-ph/0406195).

\bibitem[Bond,Pogosyan \& Souradeep 1998,2000]{bps} Bond, J. R.,
  Pogosyan, D.  \& Souradeep,T. 1998, Class. Quant.  Grav. {\bf 15},
  2671; {\it ibid.} 2000, Phys.  Rev. {\bf D 62},043005;2000, Phys.
  Rev. {\bf D 62},043006.

\bibitem[Chen et al. 2004]{gang} Chen, G.,  Mukherjee, P., Kahniashvili, T.,
Ratra, B., Wang, Y., Astrophys.J. 611 (2004) 655.


\bibitem[Coles 2003]{col03} Coles, P.  et al. 2003, {\it preprint}
(astro-ph/0310252).

\bibitem[Copi et al. 2004]{cop04} Copi, C. J., Huterer, D. \&
Starkman, G. D. 2004, Phys. Rev. D. {\it in press},
(astro-ph/0310511).

\bibitem[Cornish,Spergel \& Starkman 1998]{circles} Cornish, N.J.,
Spergel, D.N.  \& Starkman, G. D. 1998, Class. Quantum Grav., {\bf
15}, 2657

\bibitem[Cornish et al. 2003]{circles04} Cornish, N. J., Spergel, D.,
 Starkman, G. , Komatsu, E.,  2004, Phys.Rev.Lett. 92, 201302.

\bibitem[Cruz et al. 2004]{cruz}  Cruz, M., Martinez-Gonzalez, E., Vielva, P., Cayon, L., {\it preprint} (astro-ph/0405341).

\bibitem[de Oliveira-Costa et al. 2003]{angelwmap} de Oliveira-Costa,~A.,
 Tegmark,~M., Zaldarriaga,~M.  \& Hamilton,~A.  2004, Phys. Rev.{\bf D69}, 063516.

\bibitem[de Oliveira-Costa et al. 1996]{staro} de
Oliveira-Costa,~A.  Smoot, G. F.,  Starobinsky, A. A., 1996,  ApJ {\bf 468},
457.

\bibitem[Dineen et al. 2004] {coles-graca}  Dineen, P. , Rocha, G., Coles, P, {\it preprint} (astro-ph/0404356).

\bibitem[Donoghue et al. 2004]{donoghue} Donoghue, E. P.,  and Donoghue, J. F., {\it preprint} (astro-ph/0411237).

\bibitem[Durrer et al. 1998]{DKY} Durrer, R., Kahniashvili, T.  and
Yates, A.,  1998, Phys. Rev.  {\bf D 58}, 3004.

\bibitem[Ellis 1971]{ell71} Ellis, G. F. R. 1971, Gen. Rel. Grav. {\bf 2}, 7.

\bibitem[Eriksen et al. 2004a]{erik04a} Eriksen, H. K.  et al., 2004,
  Astrophys. J {\bf 605}, 14.

\bibitem[Eriksen et al. 2004b]{erik04b} Eriksen, H. K.  et al., 2004,  Astrophys. J. {\bf 612}, 64.

\bibitem[Eriksen et al. 2004c]{erik04c} Eriksen, H. K.  et al., 2004,  Astrophys. J. {\bf 612}, 633.

\bibitem[Gaztanaga \& Wagg 2003]{gaz_wag03} Gaztanaga, E. \& Wagg,
J. 2003, Phys. Rev.  {\bf D68} 021302.

\bibitem[G\'orski,Hivon \& Wandelt 1999]{hpix} G\'orski,K. M., Hivon,
E., Wandelt, B. D. 1999, in "Evolution of Large-Scale Structure",
eds. A.J. Banday, R.S. Sheth and L. Da Costa, PrintPartners Ipskamp,
NL, pp. 37-42 (also astro-ph/9812350).

\bibitem[Guth \& Pi 1982]{gut_pi82} Guth, A. H. \& Pi, S.-Y. 1982,
Phys. Rev. Lett., 49, 1110.

\bibitem[Hajian \& Souradeep 2003a]{us_prl} Hajian, A.  \& Souradeep,
T., 2003a {\it preprint} (astro-ph/0301590).

\bibitem[Hajian \& Souradeep 2003b]{us_apjl} Hajian,~A. and Souradeep,~T.,
2003b, ApJ 597, L5 (2003).

\bibitem[Hajian \& Souradeep 2004]{us_prd} Hajian, A.  \& Souradeep,
T., 2005, ApJ 618, L63. 

\bibitem[Hajian et al. 2004]{us_dodeca} Hajian,~A., Pogosyan, D. 
Souradeep,~T., Contaldi, C., Bond, R.,  2004 {\it in preparation}; Proc. 20th IAP Colloquium on Cosmic Microwave Background physics and observation, 2004. 

\bibitem[Hajian et al. 2004b]{us_magneticfield}  Hajian,~A., Chen, G., Souradeep, T., Kahniashvilli, T., Ratra, B.,  2004 {\it in preparation}.

\bibitem[Hansen et al. 2004]{han04} Hansen, F. K., Banday, A. J. \&
Gorski, K. M. 2004, {\it preprint}. (astro-ph/0404206)

\bibitem[Hinshaw et al. 2003]{hin_wmap03} G. Hinshaw, Astrophys. J. Suppl.,(2003) ,{\bf 148}, 135.

\bibitem[Komatsu et al. 2003]{kom03} Komatsu, E.  et.al., 2003, ApJS,
{\bf 148}, 119.

\bibitem[Kogut et al. 2003]{kogut_wmap03} Kogut A. et al., 2003,  Astrophys. J. Suppl.,{\bf 148}, 161.

\bibitem[Lachieze-Rey \& Luminet 1995]{lac_lum95} Lachieze-Rey, M.  and
 Luminet,J. -P. 1995, Phys. Rep.  {\bf 25}, 136.

\bibitem[Larson \& Wandelt 2004]{lar_wan04} Larson, D. L. \& Wandelt,
B. D. 2004,  {\it preprint}, (astro-ph/0404037).

\bibitem[Levin 2002]{lev02} Levin, J. 2002, Phys. Rep. {\bf 365}, 251.

\bibitem[Levin et al. 1998] {levin98}  Levin J.,  Scannapieco E., Silk J.,  (1998), Class.Quant.Grav. 15, 2689.

\bibitem[Linde 2004]{linde} Linde, A.,  (2004), JCAP 0410,  004, (astro-ph/0408164).

\bibitem[Luminet et al. 2003]{lum03} Luminet, J.-P. et al. 2003, Nature 425 593.

\bibitem[Ma \& Bertschinger 1995] {Ma95} Ma, C.-P., \& Bertschinger, E. 1995, ApJ, 455, 7

\bibitem[Mitra et al. 2004]{beam} Mitra, S., Sengupta, A., S., Souradeep, T.,{\it preprint} astro-ph/0405406

\bibitem[Munshi et al. 1995]{Munshi95}  Munshi, D., Souradeep, T., and Starobinsky, A. A., (1995), Astrophys. J., 454, 552.

\bibitem[Page et al. 2003]{page_wmap03} Page L. et al., 2003, Astrophys. J. Suppl.,{\bf  148}, 233.

\bibitem[Naselsky et al. 2004]{nas04} Naselsky et al, 2004, {\it preprint}, (astro-ph/0405523); {\it ibid} (astro-ph/0405181), 2003; ApJ,599,L53 and references therein.

\bibitem[Park 2004]{par04} Park, C., 2004, Mon. Not. Roy. Astron. Soc. {\bf
349}, 313.

\bibitem[Peiris et al., 2003]{peiris_wmap03} Peiris H.V. et al., 2003, Astrophys. J. Suppl.,{\bf 148}, 213.

\bibitem[Prunet et al. 2004]{prunet} Prunet, S.,  Uzan, J., Bernardeau, F., Brunier, T., 2004, {\it preprint} (astro-ph/0406364).

\bibitem[Ratra 1992]{ratra1992} Ratra, B.,  (1992), Astrophys.J. 391, L1.

\bibitem[Sachs and Wolfe, 1967]{SachsWolfe} Sachs, R. K., and Wolfe, A. M. 1967, ApJ, 147, 73.

\bibitem[Schwarz et al. 2004]{schw04} Schwarz, D. J. et al., 2004, {\it
preprint} (astro-ph/0403353).

\bibitem[Shafieloo \& Souradeep 2004]{shaf_sour04} Shafieloo, A. \&
Souradeep, T., 2004, Phys. Rev. D,  {\it in press}, (astro-ph/0312174).

\bibitem[Souradeep 2000]{sour00}  Souradeep,T. 2000, in `The Universe',
  eds. Dadhich, N.  \& Kembhavi, A.,  Kluwer.

\bibitem[Souradeep \& Hajian 2003]{us_pascos} Souradeep,~T. and
Hajian,~A., 2004, Pramana, {\bf 62}, 793.

\bibitem[Spergel et al. 2003]{sper_wmap03} Spergel, D. et al., 2003,
Astrophys. J. Suppl., {\bf 148}, 175.

\bibitem[Spergel et al. 1999]{Spergel98} Spergel, D., and Goldberg, D. M.,  (1999), Phys.Rev. D59, 103001.

\bibitem[Starkman 1998]{stark98} Starkman, G.  Class. 1998, Quantum
  Grav. {\bf 15}, 2529.

\bibitem[Starobinsky 1982]{star82} Starobinsky, A. A. 1982,
Phys. Lett, 117B, 175.

\bibitem[Tegmark et al. 2004]{maxwmap} Tegmark, M., de Oliveira-Costa,
A. \& Hamilton, A., 2004, Phys.Rev. {\bf D68} 123523.

\bibitem[Varshalovich et al. 1988]{Var} Varshalovich, D. A., Moskalev,
  A. N., Khersonskii, V. K., 1988 {\it Quantum Theory of Angular
  Momentum} (World Scientific).

\bibitem[Vielva et al. 2003]{vielva} Vielva, P.,  Martinez-Gonzalez, E.,  Barreiro, R. B., Sanz, J. L.,  Cayon, L., 2003, {\it preprint} (astro-ph/0310273).

\bibitem[Zaldarriaga et al. 1998]{zaldar98} Zaldarriaga, M., Seljak, U., Bertschinger, E., 1998, Astrophys. J., {\bf 494}, 491. 

\end{thebibliography}
\end{document}